\documentclass[manuscript, screen, acmsmall]{acmart}
\usepackage{tabularx}
\usepackage{multirow}
\usepackage{multicol}
\usepackage[autostyle]{csquotes}
\usepackage{array, booktabs, tabularx,makecell}

\usepackage{float}

\AtBeginDocument{%
  \providecommand\BibTeX{{%
    \normalfont B\kern-0.5em{\scshape i\kern-0.25em b}\kern-0.8em\TeX}}}

\setcopyright{acmlicensed}
\copyrightyear{2024}
\acmYear{2024}
\acmDOI{XXXXXXX.XXXXXXX}

\acmConference[CSCW '25]{Conference on Computer-Supported Cooperative Work & Social Computing (CSCW)}{October 18-22, 2025}{Bergen, Norway}
\acmISBN{XXX-X-XXXX-XXXX-X/XX/XX}

\begin{document}
\title[Photo-based Reminiscence with Older Adults]{Understanding and Co-designing Photo-based Reminiscence with Older Adults}

\author{Zhongyue Zhang}
\orcid{0009-0001-1232-004X}
\email{zzhang837@connect.hkust-gz.edu.cn}
\affiliation{
  \institution{The Hong Kong University of Science and Technology (Guangzhou)}
  \country{China}
}

\author{Lina Xu}
\orcid{0009-0003-1149-2071}
\email{xu582@connect.hkust-gz.edu.cn}
\affiliation{
  \institution{The Hong Kong University of Science and Technology (Guangzhou)}
  \country{China}
}

\author{Xingkai Wang}
\orcid{0009-0001-0220-6235}
\email{wang467@connect.hkust-gz.edu.cn}
\affiliation{
  \institution{The Hong Kong University of Science and Technology (Guangzhou)}
  \country{China}
}

\author{Xu Zhang}
\orcid{0000-0001-7937-3827}
\email{xuzhang@hkust-gz.edu.cn}
\affiliation{
  \institution{The Hong Kong University of Science and Technology (Guangzhou)}
  \country{China}
}

\author{Mingming Fan}
\authornote{Corresponding Author}
\orcid{0000-0002-0356-4712}
\email{mingmingfan@ust.hk}
\affiliation{
  \institution{The Hong Kong University of Science and Technology (Guangzhou)}
  \country{China}
}
\affiliation{
  \institution{The Hong Kong University of Science and Technology}
  \country{China}
}

\renewcommand{\shortauthors}{Anonymous Author, et al.}

\begin{abstract}
Reminiscence, the act of revisiting past memories, is crucial for self-reflection and social interaction, significantly enhancing psychological well-being, life satisfaction, and self-identity among older adults. In HCI and CSCW, there is growing interest in leveraging technology to support reminiscence for older adults. However, understanding how older adults actively use technologies for realistic and practical reminiscence in their daily lives remains limited. This paper addresses this gap by providing an in-depth, empirical understanding of technology-mediated, photo-based reminiscence among older adults. Through a two-part study involving 20 older adults, we conducted semi-structured interviews and co-design sessions to explore their use and vision of digital technologies for photo-based reminiscence activities. Based on these insights, we propose design implications to make future reminiscence technologies more accessible and empowering for older adults.
\end{abstract}




\received{16 January 2024}
\received[revised]{16 July 2024}

\maketitle

\section{Introduction}
Reminiscence, the act of revisiting past memories, is crucial for self-reflection and social interaction. It significantly enhances psychological well-being, life satisfaction, and self-identity, especially among older adults \cite{tam_effectiveness_2021, xu_effects_2023, westerhof_celebrating_2014}. Photos, with their strong visual connection to autobiographical memories, serve as powerful triggers for reminiscence \cite{conway_episodic_2009, greenberg_visual_2005, rubin_basic-systems_2005, el_haj_picture_2020}. They also function as conversational anchors, facilitating social activities and storytelling through photo sharing \cite{sit_digital_2005}. Reminiscence and related activities are crucial for older adults, enabling them to reflect on the past and connect with a broader family narrative \cite{lindley_before_2012}.

In the fields of HCI and CSCW, there is growing interest in using technology to support reminiscence for older adults. Researchers have developed and tested various reminiscence technologies aimed at making these activities more enjoyable and meaningful, while also enhancing cognitive health, engagement, and social participation among older adults (e.g., \cite{tang_memory_2007, lee_picgo_2014, piper_audio-enhanced_2013, li_exploring_2023, baker_schools_2021}). These technologies often use photos to prompt memories and facilitate storytelling and social interactions, with features such as photo annotation \cite{lee_picgo_2014} and sharing \cite{baker_schools_2021}. While promising, most of these technologies are still in the prototype stage, with evaluations based on short-term testing rather than long-term real-world use. Research on how older adults use technologies for reminiscence in their daily lives is limited, which is essential for designing solutions that meet the real-life needs and scenarios of older adults.

As photo interactions increasingly move into digital spaces, the prevalence of digital photo tools suggests that reminiscence with photos is experiencing a digital shift \cite{keightley_technologies_2014}. This shift creates opportunities to study how older adults use digital tools like digital albums and social media platforms as reminiscence technologies in daily life. Prior work suggests that this transition poses challenges, as older adults may have different understandings of digital photo operations compared to younger generations \cite{axtell_back_2019}. Despite the widespread availability of digital photo tools, their practical use and inclusivity for older adults remain underexplored, which can lead to inconsistent and unimaginative solutions \cite{axtell_back_2019}. Additionally, emerging platforms, such as short video applications, offer new opportunities for older adults to create content, fostering new forms of sharing and social interaction \cite{tang_towards_2023}. However, the specific challenges and opportunities these tools present for comprehensive reminiscence practices are not yet fully understood. Gaining a broader understanding is crucial for designing effective photo tools and interactions, as well as innovative reminiscence technologies. Addressing these gaps requires a perspective that considers how older adults use photos in their daily lives for reminiscence, beyond just storytelling or social activities in research settings.

This paper aims to fill existing gaps by providing an in-depth, empirical understanding of technology-mediated, photo-based reminiscence among older adults. We conducted a two-part user study with 20 older adults in China. In the first part, we used semi-structured interviews to explore how they currently use personal photos for reminiscence in daily life. We mapped their digital photo activities and identified their challenges and unmet needs with digital tools using the PhotoUse model, which outlines modern photo activities \cite{broekhuijsen_photowork_2017}. In the second part, we held co-design sessions to explore their ideal photo-based reminiscence experiences mediated by technologies.

Our findings indicate that realistic photo-based activities significantly support reminiscence among older adults. However, current digital tools often fall short in providing adequate support, particularly for tasks such as annotating digital photos. We also identified new opportunities introduced by emerging tools, such as those that facilitate the creation of videos with old photos, fulfilling needs for memory authoring and creative expression. These insights go beyond prototype designs to address real-life needs and scenarios. Additionally, our co-design sessions uncovered various creative ideas from participants, such as co-creating photo artifacts with friends by attaching voice to photos. Based on these findings, our research suggests that reminiscence technologies for older adults should be accessible, sensitive, connected, creative, and culturally aware, offering novel directions and implications for future reminiscence technologies.

The main contribution of this work lies in the empirical insights we present on how older adults use and envision digital tools for practical and realistic photo-based reminiscence activities. These insights lead to design implications aimed at enhancing technologies that support reminiscence among older adults through photo-related activities. This work extends prior research on reminiscence technologies and photo-related tools for older adults and is broadly applicable to understanding reminiscence in photo-related activities.

\section{RELATED WORK}
\subsection{Photo-based Reminiscence of Older Adults}
Reminiscence, which involves revisiting memories of past events, is an effective means for enhancing psychological well-being, life satisfaction, self-esteem, and mood in older adults, including those with and without cognitive impairments such as dementia \cite{tam_effectiveness_2021, xu_effects_2023}. It serves three key functions: \textit{social functions}, fostering bonding between people through the sharing of personal memories; \textit{instrumental functions}, recalling coping strategies for current challenges and facilitating emotional regulation; and \textit{integrative functions}, aiding in identity adjustment by reflecting on the past \cite{westerhof_celebrating_2014}. This activity is crucial for older adults, contributing to personal reflection and family narratives. It also facilitates connections with present family members through the creation of artifacts like mementos, enabling older adults to leave a legacy \cite{lindley_before_2012}.

Reminiscence is part of the autobiographical memory system, encompassing the encoding, storage, and retrieval of episodic information \cite{bluck_reminiscence_1998}. It typically involves cues such as verbal prompts, photos, or music to trigger autobiographical memories. In everyday situations, older adults often find storytelling enjoyable and identity-enhancing, especially when prompted by photos or verbal cues, helping to reduce social isolation \cite{thiry_unearthing_2012}, an increasingly serious issue for this demographic \cite{committee_on_the_health_and_medical_dimensions_of_social_isolation_and_loneliness_in_older_adults_social_2020}.

Photos are particularly effective in triggering reminiscence, as autobiographical memories are often constructed as visual images. This makes photos a powerful medium for memory retrieval \cite{conway_episodic_2009, greenberg_visual_2005, rubin_basic-systems_2005}. They are more effective than verbal prompts due to their richness in visual information, which facilitates the memory retrieval process and enhances the visual construction of memories \cite{el_haj_picture_2020}. The inherent nature of photos, serving as conversational anchors \cite{sit_digital_2005}, aligns with the social aspects of reminiscence. This facilitates social activities like narrative storytelling and memory sharing, thereby strengthening connections with family and friends \cite{keightley_technologies_2014}, and fostering inter-generational interaction \cite{kang_momentmeld_2021}.

Acknowledging the importance of reminiscence for older adults and the benefits of photos in facilitating this process, our work delves deeper into the role of photos in the lives of older adults. We provide an empirical understanding of how older adults in China use personal photos for reminiscing in their daily lives and seek ways to support this practice.

\subsection{Technology-mediated Reminiscence for Older Adults}
Technology is increasingly recognized for its role in enhancing reminiscence activities for older adults. Advances in multimedia resources are essential for reminiscence therapy, offering benefits such as improved conversational control, adaptation to motor deficits, and enhanced social engagement for dementia patients \cite{lazar_systematic_2014}. The HCI and CSCW communities focus on designing technology-mediated reminiscence activities that promote healthy aging.

A key objective in this area is to improve memory recall and exercise in older adults by providing meaningful experiences and memory cues through technology. For example, Memory Karaoke uses contextual information like photos, time, and location to reinforce memory exercises \cite{tang_memory_2007}. Tools like Picgo, which allow for photo annotation, help caregivers guide older adults through memory exercises \cite{lee_picgo_2014}. Voice-activated devices offering positive reminiscence prompts have been shown to enhance well-being \cite{cuadra_designing_2023}, while robot-assisted reminiscence, where robots use photos to start conversations, creates unique engagement opportunities \cite{wu_interactive_2020}. Using photos as memory cues is crucial for enriching memory experiences \cite{lee_providing_2007}.

Research also aims to foster social interactions among older adults, their families, and caregivers to reduce social isolation \cite{piper_audio-enhanced_2013, li_exploring_2023, baker_schools_2021}. Technologies like VR and AR create immersive environments for remote social reminiscence \cite{baker_schools_2021}, enable intergenerational storytelling through personal photos \cite{li_exploring_2023}, and match semantically related photos to enhance empathy across generations \cite{kang_momentmeld_2021}. To encourage older adults' participation in the online community, Pinteresce is a prototype designed to leverage family-generated reminiscence prompts on Pinterest, simplifying the interface to foster photo sharing and commenting \cite{brewer_pinteresce_2015}. This promotes social interaction and reduces isolation by fostering meaningful connections with family and friends. In these scenarios, photos serve as both memory aids and conversation starters.

These studies demonstrate the potential of technology-mediated reminiscence. However, most current technologies are still in the prototype phase and have only undergone short-term testing. Although the 5-month deployment study of a paper-digital photo album provides insights in effectively engaging older adults' interests and encouraging social interaction \cite{piper_audio-enhanced_2013}, a comprehensive understanding of how older adults use technologies in their daily lives for reminiscence, the technology-mediated reminiscence activities they engage in, and their needs remains limited. Given the complex and dynamic nature of reminiscence \cite{webster_mapping_2010}, it is essential to gain a deep, empirical understanding of technology-mediated reminiscence. Our study aims to fill this gap by examining the practical experiences and challenges older adults face when using technologies for daily reminiscence, specifically with photos.

Additionally, we draw from research suggesting that memory tools should support creativity and active meaning-building in older adults, rather than treating them as passive recipients of memory cues \cite{petrelli_making_2009, thiry_authoring_2013}. While memory triggers can help older adults revisit memories and stay cognitively active, understanding their preferred activities and design preferences for reminiscence technologies is crucial. Therefore, we used a participatory design approach in our study to explore engaging and meaningful technological designs \cite{davidson_participatory_2013}. We used existing digital photo interactions as inspiration, encouraging older adults to imagine their ideal photo-based reminiscence experiences. Our findings indicate that reminiscence technologies for older adults should be accessible, sensitive, connected, creative, and culturally aware. By aligning future designs with their reminiscence needs and preferences, we aim to make technology-mediated reminiscence more meaningful and empowering for older adults.

\subsection{Digital Photo Interactions and the PhotoUse Model}
To understand how people interact with photos, various studies have defined the activities and methods associated with personal photo use. Frohlich et al. identified the requirements for digital photoware based on real family use cases, emphasizing photo-sharing practices as ways to review and communicate experiences with others \cite{frohlich_requirements_2002}. Further research explored the concept of PhotoWork, identifying three phases of working with digital photos: pre-download, at-download, and pre-share \cite{kirk_understanding_2006}. Keightley et al. later examined the digital shift of photography in remembering practices, categorizing activities into photo-taking, photo-storing, photo-viewing, and photo-sharing \cite{keightley_technologies_2014}. A follow-up study of PhotoWork introduced the PhotoUse model, categorizing photo activities into four broad areas: accumulating, curating, retrieving, and appropriating photos \cite{broekhuijsen_photowork_2017}. While mirroring the four categories identified by Keightley et al., this model provided a more detailed list of specific activities within each category, suggesting that photo activities are multithreaded and interactive processes without a distinct start and finish. It highlighted how activities such as editing, triaging, collecting, organizing, and viewing photos often combine towards reminiscence goals, such as sharing, storytelling, and self-presentation. However, it did not delve into the details of photo activities related to reminiscence and their support by digital tools.

These studies primarily illustrate general photo activities for a broad audience without specifically targeting older adults. Among these models, the PhotoUse model is relatively more comprehensive and detailed, enabling researchers to systematically analyze specific applications and assess how they meet different photo activity needs. For instance, Axtell et al.'s research mapped eight photo-related tools (e.g., Instagram, Apple iCloud, Dropbox) to the activities within the PhotoUse framework, helping to understand the breadth of tasks involved in digital photo interactions and allowing for comparison and analysis of these tools \cite{axtell_underdeveloped_2023}.

Due to its relative comprehensiveness, detail, and applicability, we decided to apply the PhotoUse model as a way to understand the photo-based reminiscence practices of older adults. Instead of analyzing the functionalities existing tools provide, as in Axtell et al.'s work, we examine which activities are involved when reminiscing with photos and whether tools support these activities.  Older adults have been found to have different mental models of photo storage applications compared to younger generations, lacking a consistent and recognizable process and feeling pressured to learn more about the tools they use \cite{axtell_back_2019}. As digital photos become prevalent for photo-based reminiscence, it is crucial to determine whether the design of these photo-based operations is accessible and inclusive for older adults when they conduct realistic photo-based reminiscence.

This systematic approach offers a new perspective on understanding older adults' needs for photo-based reminiscence, which is crucial for enhancing the effectiveness of digital tools and technologies designed to support this activity.

\section{METHOD}

Our method involved a two-part user study. The first part consisted of semi-structured interviews to understand the practices of photo-based reminiscence among older adults, as well as the challenges they face when using related digital tools. This was followed by a co-design session aimed at exploring their design preferences.

\subsection{Participants}
We recruited 20 older adults (gender: 14 female, 6 male; age: median = 62, mean = 63.65, SD = 6.48) for our study via social media platforms and snowball sampling. Participants were recruited from an urban community in China, with an inclusion criterion of age 55 and above, aligning with the retirement age of civil servants in China  \cite{wikipedia_contributors_retirement_2024}. Before participating, all individuals provided informed consent, agreeing to partake either individually or in self-formed small groups of families or friends. We conducted a total of 12 sessions (see \autoref{tab:commands}), each comprising 1-3 participants and lasting approximately 60-90 minutes. All participants were Chinese, and the user studies were conducted in Mandarin. Participants were compensated in accordance with local standards.

\begin{table*}
\small
  \caption{Demographic of participants (N=20)}
  \label{tab:commands}
  \begin{tabular}{lllll}
    \toprule
Session ID & Participant ID & Age & Gender & Relationship\\
    \midrule
    \multirow{2}{*}{1}& \multicolumn{1}{l}{P1} &\multicolumn{1}{l}{73} & \multicolumn{1}{l}{F} & \multirow{2}{*}{Couple}\\
            & \multicolumn{1}{l}{P2} &\multicolumn{1}{l}{73} & \multicolumn{1}{l}{M}\\
            \cline{1-5}
    \multirow{3}{*}{2} & \multicolumn{1}{l}{P3} &\multicolumn{1}{l}{58} & \multicolumn{1}{l}{F} & \multirow{3}{*}{Siblings: P3 P5; Couple: P4 P5}\\
            & \multicolumn{1}{l}{P4} &\multicolumn{1}{l}{80} & \multicolumn{1}{l}{M}\\
             & \multicolumn{1}{l}{P5} &\multicolumn{1}{l}{63} & \multicolumn{1}{l}{F}\\
             \cline{1-5}

            3 & P6 & 62 & F & /\\
            \cline{1-5}
            4 & P7 & 68 & M & /\\
                \cline{1-5}
            \multirow{2}{*}{5}& \multicolumn{1}{l}{P8} &\multicolumn{1}{l}{58} & \multicolumn{1}{l}{F} & \multirow{2}{*}{Couple}\\
            & \multicolumn{1}{l}{P9} &\multicolumn{1}{l}{58} & \multicolumn{1}{l}{M}\\
            \cline{1-5}
 \multirow{3}{*}{6} & \multicolumn{1}{l}{P10} &\multicolumn{1}{l}{66} & \multicolumn{1}{l}{F} & \multirow{3}{*}{Friends}\\
            & \multicolumn{1}{l}{P11} &\multicolumn{1}{l}{69} & \multicolumn{1}{l}{F}\\
             & \multicolumn{1}{l}{P12} &\multicolumn{1}{l}{60} & \multicolumn{1}{l}{F}\\
             \cline{1-5}
    \multirow{2}{*}{7}& \multicolumn{1}{l}{P13} &\multicolumn{1}{l}{62} & \multicolumn{1}{l}{F} & \multirow{2}{*}{Couple}\\
            & \multicolumn{1}{l}{P14} &\multicolumn{1}{l}{66} & \multicolumn{1}{l}{M}\\
            \cline{1-5}
            8 & P15 & 55 & F & /\\
            \cline{1-5}
            9 & P16 & 58 & F & /\\
            \cline{1-5}
            \multirow{2}{*}{10}& \multicolumn{1}{l}{P17} &\multicolumn{1}{l}{61} & \multicolumn{1}{l}{F} & \multirow{2}{*}{Couple}\\
            & \multicolumn{1}{l}{P18} &\multicolumn{1}{l}{57} & \multicolumn{1}{l}{M}\\
            \cline{1-5}
            11 & P19 & 59 & F & /\\
            \cline{1-5}
            12 & P20 & 66 & F & /\\
           
    \bottomrule
  \end{tabular}
\end{table*}

\subsection{Preparation}
To align with our research objectives, we asked participants to bring old personal photos, typically black-and-white, that they frequently use for reminiscing in their daily lives. These photos, whether physical or digital, were expected to hold personal significance or unique value for the participants, enhancing the depth and impact of our discussions \cite{lee_providing_2007}. We emphasized selecting images that participants felt comfortable sharing to mitigate privacy concerns and avoid unintentionally triggering distressing memories.

\begin{figure}
    \centering
    \includegraphics[width=1\linewidth]{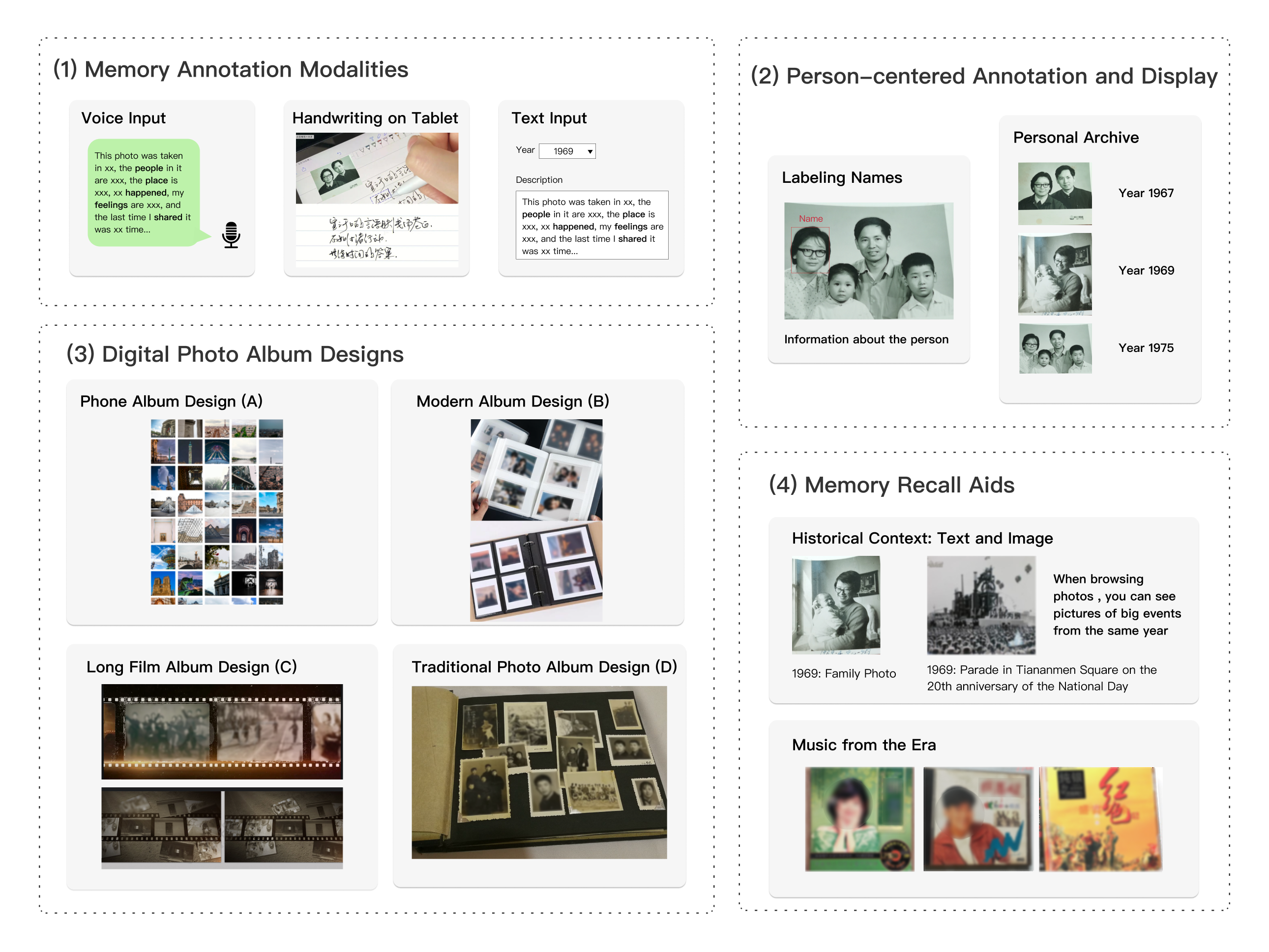}
    \caption{In our participatory design sessions, we used 11 discussion cards, categorized into four groups: 1) Memory Annotation Modalities (text in the voice input and "Description" box: "This photo was taken in xx, the people in it are xxx, the place is xxx, xx happened, my feelings are xxx, and the last time I shared it was xx time ..."), 2) Person-Centered Annotation and Display, 3) Digital Photo Album Designs, and 4) Memory Recall Aids. We have obtained the permissions to use personal black and white photos featured in the cards.}
    \label{fig:cards-overview}
\end{figure}

In our co-design sessions, we adopted a participatory design approach centered on critiquing and discussing design concepts proposed by the research team. Eleven discussion cards (Figure \ref{fig:cards-overview}), created based on insights from relevant literature and findings from a pilot study (see Appendix \ref{appendix: pilot}), represent core digital interactions in photo-based reminiscence, such as annotation, retrieval, and display of photos. These categories were purposefully chosen to ground the reminiscence process in familiar activities, while also suggesting digital enhancements. Recognizing that older adults may be less inclined to engage in sketching activities and may find abstract technical concepts challenging to interpret \cite{xie_connecting_2012}, each card was designed to be visually accessible and conceptually concrete. Rather than acting as prescriptive prototypes, these cards served as prompts to encourage dialogue and creativity, providing older adults with a clear foundation to express preferences and offer feedback.

\begin{enumerate}
\item \textbf{Memory Annotation Modalities}: 
This category explores diverse ways to annotate photos, drawing from research on the role of annotations in enhancing memory recall \cite{lee_picgo_2014,lee_picmemory_2016}, storytelling \cite{axtell_photoflow_2019} and retrieval \cite{rodden_how_2003}. By offering multiple input methods (e.g., voice, handwriting, text), this card set allows participants to reflect on which types of annotation might best support their memory and storytelling preferences, while accommodating varying levels of digital literacy and accessibility needs.
\item \textbf{Person-centered Annotation and Display}: 
This category draws from pilot study insights and prior research on the importance of significant individuals in older adults’ photo collections and their storytelling tendencies to emphasize people and chronological changes \cite{li_exploring_2023}. By prioritizing tagging and organizing photos based on important individuals—similar to systems like Kim et al.’s personal photo diary \cite{kim_photo_2011}—these cards are designed to encourage discussions on person-centered retrieval, facilitating easier recall and organization.
\item \textbf{Digital Photo Album Designs}: Drawing from findings on the importance of visual organization in photo display on memory evocation and social reminiscence \cite{axtell_design_2022,axtell_underdeveloped_2023}, this category introduces a variety of digital album layouts to elicit preferences for viewing and interacting with digital photo collections. This category aims to help older adults express preferences in how they would ideally access and share these photos digitally.
\item \textbf{Memory Recall Aids}: Research shows that while photos activate episodic memory, additional cues can enhance this \cite{tang_memory_2007} including semantically similar photos \cite{kang_momentmeld_2021}, broader visual contexts \cite{broekhuijsen_design_2017}, and augmented elements \cite{li_exploring_2023}. Our pilot study found older adults' interest in historical context for their photos, so we used a discussion card pairing personal photos with relevant historical images to gauge interest in contextual digital photo tools. Recognizing music's role in enriching reminiscence \cite{istvandity_combining_2017, kim_slide2remember_2022}, we also introduced a music-focused card to explore integrating multi-sensory elements in reminiscence activities.
\end{enumerate}

\subsection{Process}
The study was conducted in a conference room at our institution, which was suitable for face-to-face interactions with participants. Initially, we presented an overview of the research context and session structure, addressing questions and concerns. Consent for audio recordings was obtained before each session. In every session, two to three researchers were involved: one leading the session, while the others observed, took notes, and provided assistance as needed.

\begin{figure}
    \centering
    \includegraphics[width=0.8\linewidth]{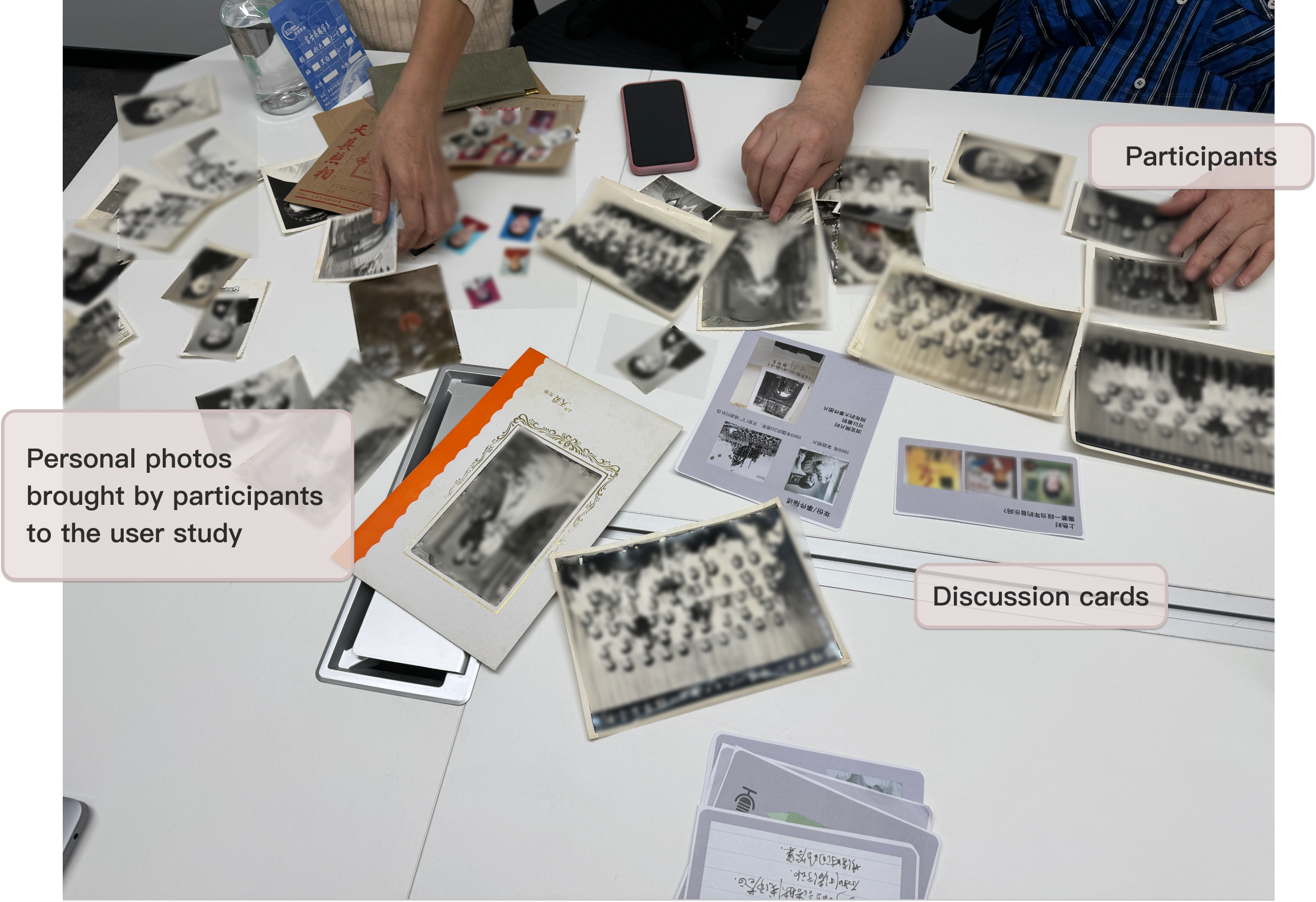}
    \caption{Photo captured during the user study with two participants} (with blur effect on participants' personal photos).
    \label{fig:userstudy}
\end{figure}

\subsubsection{Part One: Semi-structured Interview}
This session aimed to explore older adults' engagement with personal photos during reminiscence activities and identify their needs and challenges with digital tools. Each interview began with a 5-minute informal chat about the participants' backgrounds and hobbies to build rapport. We then posed guided questions about their old personal photo collections, motivations for reminiscing, activities with these photos, and emotional experiences during these activities. Examples of questions included, "How often do you look at old photos? In what scenarios?" and "What are you thinking when looking at the old photos?" Participants were encouraged to share photo-linked stories, providing details such as time, location, and people to understand their storytelling and memory recall methods. Questions included, "Please share a story from one of the photos" and "Do you encounter difficulties recalling details when reminiscing with these photos?"

If a specific photo activity listed in the PhotoUse Model was mentioned (e.g., annotating photos, which falls under the \textit{"organizing"} subcategory within \textit{"curating"}), we delved into their motivations and challenges with the activity and related tools. To ensure comprehensive coverage of photo activities that older adults engage in during reminiscence with digital tools, we followed up on unmentioned activities with deliberate questions. For instance, we asked, "Have you tried to edit your old photos?" If participants responded affirmatively, we inquired further about their methods and experiences. If they had not tried, we discussed whether they had an interest in this activity and what might prevent them from engaging in it.

\subsubsection{Part Two: Participatory Design}
This session aimed to understand older adults' preferences for photo-based reminiscence designs. After completing inquiries about all photo-related activities with the older adults, we transitioned to this part of the session. We explicitly informed the participants that we wanted to integrate their creative ideas into the design of technologies that support reminiscence with their photos. We presented design cards and encouraged them to freely comment on whether they liked the ideas, found them practical and helpful, or did not like them at all.

When presenting the ideas, we ensured that the older adults could understand each card's meaning, explaining until they grasped the core concept. Participants usually started by commenting on whether they liked or disliked each idea and provided their reasons. Since we had discussed their unmet needs for reminiscing with digital photos in detail during the previous semi-structured interview part, some participants actively drew on their personal experiences. To further facilitate conversation, we also referred back to their earlier comments. For example, we might say, "You mentioned that annotating photos feels tiring. Now, if you could annotate in several ways, which one would you prefer? (pointing and explaining Memory Annotation Modalities)" or "(After going over design cards) You mentioned that children are less aware of family traditions. Do you think anything could help with that?" The process was flexible, with discussion cards acting as a starting point to spark creativity rather than to design anything based on the cards we provided. Our primary focus was on helping older adults create designs that matched their most urgent needs for using digital tools for reminiscence. If a card did not match their interests, we did not dwell on it for long.

We provided papers and pencils on the table but assured participants that drawing was not required to encourage their participation. Participants generally preferred verbal expressions, so we actively echoed back their ideas to ensure we understood them correctly.

\subsection{Data Analysis}
We collected approximately 15 hours of audio recordings from 20 participants across 12 sessions. The data collection was supplemented by 1) quick field notes taken by researchers, 2) photos captured during interviews showcasing photos and digital tools, and 3) photos from participatory design activities showing design card preferences. The audio recordings were transcribed into text paragraphs using a commercial Automatic Speech Recognition (ASR) system, iFLYTEK \footnote{https://www.iflyrec.com/zhuanwenzi.html}, and the transcriptions were verified by our research team for accuracy.

In analyzing the data, we adopted a thematic analysis \cite{braun_using_2006} approach. The first author engaged in multiple readings of the transcripts and reviewed the supplementary materials to generate initial codes. These initial codes were then carefully developed into themes. To ensure thoroughness and consensus in our analysis, three authors collaboratively reviewed, refined, and developed these themes into sub-themes through  discussions in weekly meetings. This iterative process of theme generation and refinement was sustained throughout the analysis and writing process.

To analyze digital tool use and challenges in photo-based reminiscence, we applied the PhotoUse model \cite{broekhuijsen_photowork_2017} for mapping photo activities. The PhotoUse model categorizes photo activities into four broad areas: accumulating, curating, retrieving, and appropriating photos. Each category contains more specific activities, which are detailed in Figure \ref{fig:photouse}. For each participant, two co-authors independently coded the data by listing the specific activities involved (according to the definition provided in Appendix \ref{appendix: model}) and the relevant tools used. They then discussed their findings to refine the codes. The first author organized the results into a final chart and summarized the tool categories involved. Our research focuses less on individual behaviors and more on presenting an overview of the photo activities our participants engage in with digital tools, the categories of tools they use, and whether their needs are supported or unsupported by these tools. If a participant engaged in a specific activity with a tool, the corresponding space was marked with an "x". For these activities, we identified four challenges participants face when using these tools for reminiscence (marked as "1"-"4"). Additionally, we noted two activities that older adults engage in but are not supported by any digital tools (marked as “A”), two activities that older adults engage in but are not supported by any digital tools and are also less dependent on digital tools (marked as “B”), and two activities currently engaged with digital photos, but challenges mainly caused by preceding activities (marked as “C”). These preceding activities refer to preparatory steps, such as content triaging and searching for suitable materials needed for sharing or collaging, which can be time-consuming and effort-intensive due to inadequate support from current digital tools.

Consensus on themes was reached team-wide, and quotes were translated by the first author and verified by co-authors.

\section{Findings}
We presented our key findings of the two parts of our study: Based on part 1 of our study (i.e., semi-structured interview using old personal photos), we first showed older adult's daily practices with photo-based reminiscence, and identified the photo activities and four challenges with relevant digital tools. Based on part 2 of our study (i.e., co-design activities using discussion cards), we showed their preferences for digital tool designs for enhancing reminiscence experiences.

\subsection{Personal Photo-based Reminiscence in the Daily Lives of Older Adults}
Our interviews with participants aimed to understand how they engage in reminiscing through old personal photos. We structured our presentation around key subthemes: Engagement and Motivation, Organization, Sharing, Emotional Experience, and Storytelling Patterns.
\subsubsection{Engagement and Motivation}
In our study, the majority of participants highly valued old photos stored in physical albums, often displayed on shelves or kept in boxes. These albums served as repositories of cherished memories, featuring family, friends, classmates, colleagues, and unique instances like one participant's self-developed microscope images. However, participants experienced photo loss during relocation—some preserved only a few, while some nearly lost their entire collections. 

The act of revisiting these photos often occurred spontaneously, typically during leisurely moments or while reorganizing the home. However, participants noted a decline in the frequency of viewing physical photos as they aged, e.g.,\textit{"I seldom flip through the physical albums myself unless there's a specific reason to"} (P8). 

External events serve as catalysts for engaging with these photos. The most commonly mentioned event is group reminiscence with old classmates, which occurs both in-person reunions and through online photo-sharing chats. Other notable occasions include sharing similar or relevant memories with friends or family, encountering nostalgia-triggering events, commemorating past family members in daily life or on special occasions, and conducting storytelling sessions with grandchildren.

\subsubsection{Organization}
Among our participants, many expressed a keen interest in reorganizing their photo collections to facilitate better reminiscence activities. However, only a few had actively undertaken this task by creating new physical albums. P6 went a step further by supplementing these albums with annotated notes detailing significant yearly events, which she thought led to more frequent reminiscence and sharing. As P6 remarked,\textit{
"I share the new album whenever friends visit, and I am delighted by how much everyone appreciates it"} (P6).

Digitization emerged as a key theme, seen as a bridge to preserving memories and sharing them across generations. However, this transition from physical to digital was fraught with challenges, primarily technological barriers and a lack of guidance, e.g., \textit{"No one is here to guide us, and our learning ability is not good"} (P9). Some considered using professional digitization services but were deterred by costs and perceived complexities of communications, e.g., \textit{"I've been there several times, but I was hesitant to go in... I was just afraid that people haven't done it before and then waste time talking about it"} (P19). As a result, participants found using their mobile phones for photo preservation and interaction a more feasible option. However, only a few participants organized their old photos deliberately by creating specific albums on their phones, while most mixed them with their current digital collections.

\subsubsection{Sharing}
These photos were typically shared within close circles of family and friends or on private social media spaces. The emergence of creating videos from old photos on the video authoring tools provided by short-form video sharing platforms (SVSPs) introduced a novel way of sharing, infusing emotional depth and narrative into memories, as P20 expressed, \textit{"When I share, it's in a series, telling a story by nature. This allows viewers to understand the emotions behind them and connect more deeply"}. Participants usually use existing templates to upload a series of photos, placing significant emphasis on choosing music consistent with the feelings and themes of the photos. P16 further employed Douyin\footnote{Douyin is a short-form video sharing mobile application. It is the Chinese version of TikTok.} to leverage AI-generated tags for inspiration in crafting accompanying narratives. Although having passion for making videos, participants showed less interest in posting the videos on the platforms due to privacy concerns, some hesitating even to post on their private social media accounts, as they did not want to expose family conditions to less familiar audiences.

While all participants had photos to share, they recognized this wasn't a universal experience. P13 mentioned it is common that many of her friends had no photos to share, and participants who retained unique photos like graduation group pictures, often became the custodians of shared memories for their wider social circles.

\subsubsection{Emotional Experience}
Participants emphasized the crucial role of emotions in the reminiscence process, acknowledging that emotions often act as catalysts for engaging with memories. As P17 expressed, \textit{"I feel that when emotions are involved, that's the main time we want to reminisce. Without emotions, we probably seldom recall (these memories)}"

Our research reveals a diverse and multifaceted range of emotional responses to photos. For many participants, revisiting personal photos brought forth positive emotions such as happiness, friendship, and nostalgia. Intriguingly, some cherished reminiscing about their youthful appearances, revealing a bittersweet mix of nostalgia and self-reflection. Conversely, memories associated with lost family members often elicited sadness. P15, with a family history marked by societal upheaval, faced a dichotomy: a desire to preserve memories clashed with reluctance to revisit them. This sentiment was further complicated by uncertain fates of individuals in these photos, leading to feelings of pity and sorrow for some participants. Reminiscence with personal photos evoke both positive and negative emotions, as P4 articulated, \textit{"There's heartache, there's happiness"}.

The persistent march of time emerged as a recurrent theme, encapsulated by expressions such as \textit{"Suddenly, decades have passed"} (P8). Some participants delved into the profound impact and intricacy of their memories, e.g., \textit{"How to express this feeling? Our generation has been through a lot. It's very complicated"} (P1). P9 shared a deeply personal insight, acknowledging the dual nature of memories as both stirring and unsettling, \textit{"Looking at these old photos makes me reflect on my youth, my education, and early career. Sometimes, these thoughts are so intense that they disturb my sleep"}. 

However, P6 discovered tranquility in organizing photos, experiencing a serene reflection on life's journey, \textit{"Organizing these photos brings peace. My heart feels calm, recalling joyful times. Even photos that bring sadness, like those of my deceased sister, are cherished"}. P7 offered a more contemplative view, regarding old photos as mirrors reflecting upon life's events and personal growth, \textit{"I don't necessarily experience many emotions looking at old photos. Instead, I find myself reflecting on interesting events or areas for improvement".} Lastly, P17 highlighted the dynamic nature of emotions linked to photos, observing how they evolve with life's changes and circumstances.

\subsubsection{Storytelling Patterns} During the interviews, we encouraged older adults to share stories related to the images,  and we asked about relevant information. Our primary goal was to understand their experiences of memory recall and storytelling when interacting with old photos. We assumed that this understanding could be instrumental in designing features that enhance memory recall, particularly regarding the annotations of information. We identified and presented two storytelling patterns observed during this process:

\textbf{Deducing Details through Contextual Clues.} Our study revealed a distinctive storytelling pattern among participants when reminiscing about photos. They often employed contextual clues to deduce specific details like time, location, people involved, or colors in the photos. These cues ranged from personal life stages to historical markers, enriching their storytelling. This approach allowed them to weave objects in the photos into their broader personal narratives. 

For instance, P19 used contextual clues to pinpoint a date and reminisce about their work experience, \textit{"This (photo with colleagues) should be from around 1984, because we started working in 1980..."}. Similarly, P13 and P14 collaboratively deduced the time of a photo, linking it to significant personal milestones, \textit{"This photo was taken around our marriage... it seems like Shanghai, near the Huangpu River... around 1980 or 1982. I graduated in 1982, and we got married in 1985, so it must be 1985"}.

This characteristic underscores a critical aspect of their reminiscence process: when attempting to recall specific information from a photo, participants do not directly arrive at an answer. Instead, they rely on a deductive process during their storytelling to piece together the clues from the photo.

\textbf{Resorting to Common Memories Amidst Uncertainty.} Participants frequently used generalized memories to navigate uncertainty. Phrases like \textit{“everything back then was like this”} became a common refrain, especially when speculating about colors in old black and white photos. This tendency towards generalization was evident in statements such as \textit{“At that time, most people wore blue or black”} (P1) and \textit{“The pants then were likely green or blue”} (P3). This approach served as a coping mechanism for gaps in memory, allowing participants to fill in uncertain details with common trends of the era.

Contrastingly, when specific memories were vivid, participants eschewed generalizations in favor of detailed, personal narratives, e.g., \textit{"My clothes were brown corduroy because my mother made them"} (P5), \textit{"This dress is pure white with pink stripes, probably the prettiest I had. My aunt brought it from Hong Kong"} (P6). 

This dual tendency, as seen co-existing in some participants, highlights how individuals navigate between distinct memories and generalized recollections.

\subsection{Challenges of Digital Tools in Supporting Photo-based Reminiscence for Older Adults}

In this section, we presented general digital tool uses for photo-based reminiscence and identified four major challenges: 1)  Digitize Photos into High Quality Artifacts, 2) Long-term Preservation, 3) Photo Annotation, and 4) Photo Retrieval.

We added blur effect to all images containing personal images in this section to protect participants' privacy.

\subsubsection{Overview of Digital Tool Use for Photo-Based Reminiscence}
\begin{figure}
    \centering
    \includegraphics[width=1\linewidth]{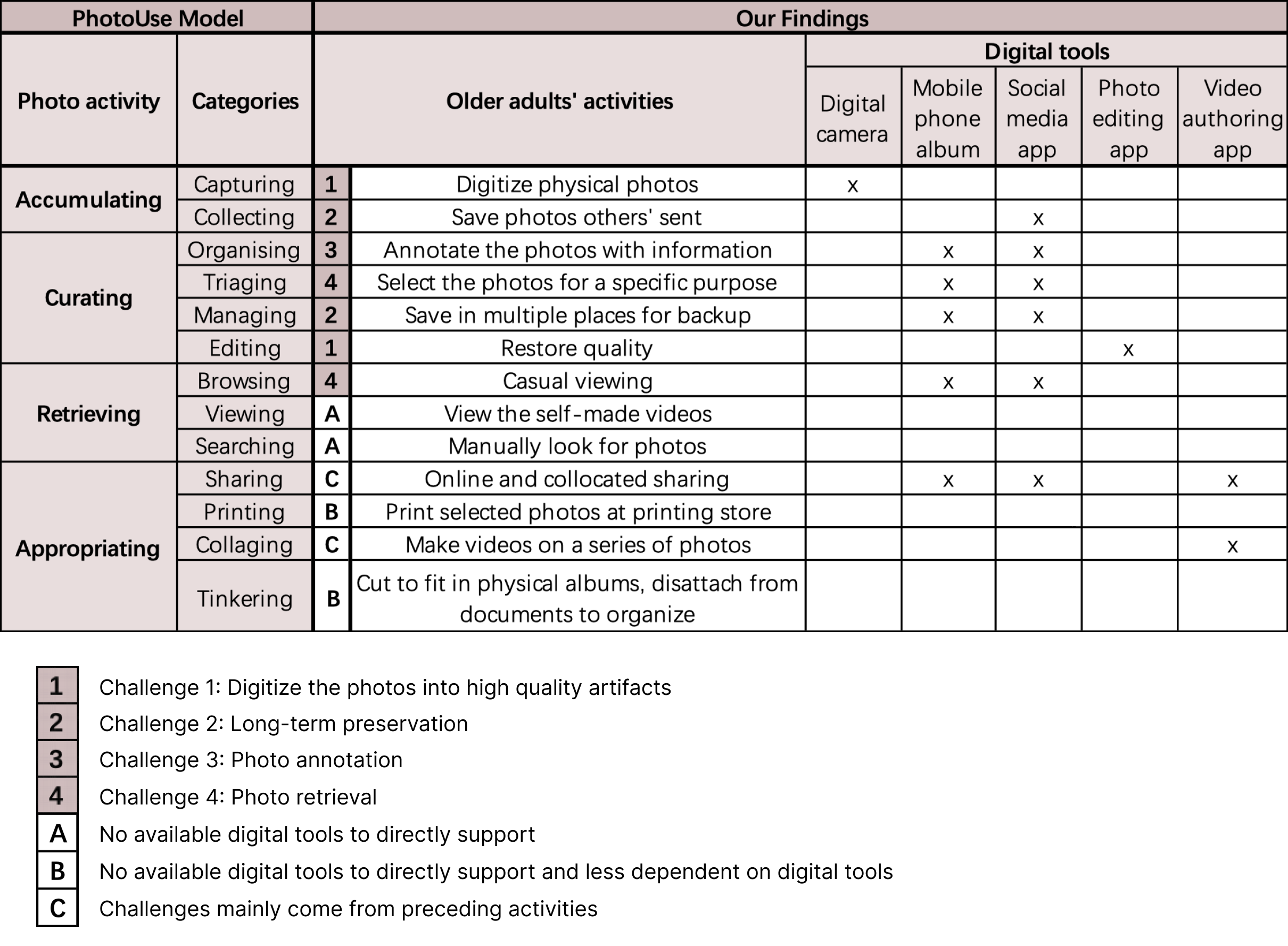}
    \caption{The overview of older adults' photo activities and their usage of digital tools as elaborated in Section 4.2. On the left side, the PhotoUse Framework \cite{broekhuijsen_photowork_2017} outlines four general photo-related activities and their specific categories. On the right side, we indicate which digital tools participants used for each activity, marked with an "x." For each activity, under "Older adults' activities", we provide specific examples mentioned in Section 4.2.}
    \label{fig:photouse}
\end{figure}
In our research, older adults engage in diverse photo activities when reminiscence with personal photos, including accumulating, curating, retrieving, and appropriating photos, as detailed in the model (see \autoref{fig:photouse}). Their primary device is the mobile phone, and they utilize various mobile applications, including digital cameras, mobile phone albums, social media platforms, photo editing, and video authoring applications.

Curating activities, such as annotating photos, along with viewing and sharing, are typically conducted within mobile phone album apps that offer diverse functionalities. Older adults also extensively use social media applications for photo-based reminiscence. These platforms cover curating, viewing, and sharing photos while facilitating additional social interactions, including collecting photos from others. Many participants reported using the advanced photo album features within these social media apps similarly to dedicated photo album apps.

Additionally, older adults use photo editing tools to enhance image quality and video authoring tools to create narrative content for viewing and sharing, which goes beyond typical photo tools for database or narrative purposes (e.g. social media platforms) \cite{axtell_back_2019}. However, photo retrieval, especially for passive viewing or targeted searching, appears to be inadequately supported by current digital tools. While creating videos for enjoyable photo viewing is common, it is effort-intensive. As P20 stated, \textit{"It's really troublesome, really tiring...I ended up working for two nights straight"}. For targeted searching, the lack of effective tools often results in reliance on manual methods. The main challenge for both photo sharing and collaging arises from preceding activities, such as content triaging and searching for suitable content for sharing and collaging activities like video making.

Printing and tinkering activities seem to be not supported and less dependent on digital tools. While many participants expressed a desire to print their photos, acknowledging its positive effect on eye comfort and appreciation, printing usually requires extra devices and assistance from others, making it less accessible. The original definition of tinkering in the PhotoUse model pertains to printed photos, so it does not readily apply to digital photos. We observed that participants often cut their original photos to paste in albums, whereas they rarely did this with printed photos. This may be attributed to the limited accessibility of printing, resulting in fewer printed photos available. Conversely, participants familiar with video making often manually 'cut' important elements from their photos to create videos. However, these videos typically focus on daily life rather than reminiscence, as participants prefer to preserve the originality of their old photos.

\subsubsection{Challenge 1: Digitize Photos into High Quality Artifacts}

 A significant need was identified among participants for effective photo digitization and quality preservation using their mobile phones, their most accessible tool. They sought improved methods for digital capture (\textit{"capturing"}) and high-quality restoration (\textit{"editing"}) of old photos which impact their motivation and sharing experiences.

\begin{figure}
\centering
\includegraphics[width=0.3\linewidth]{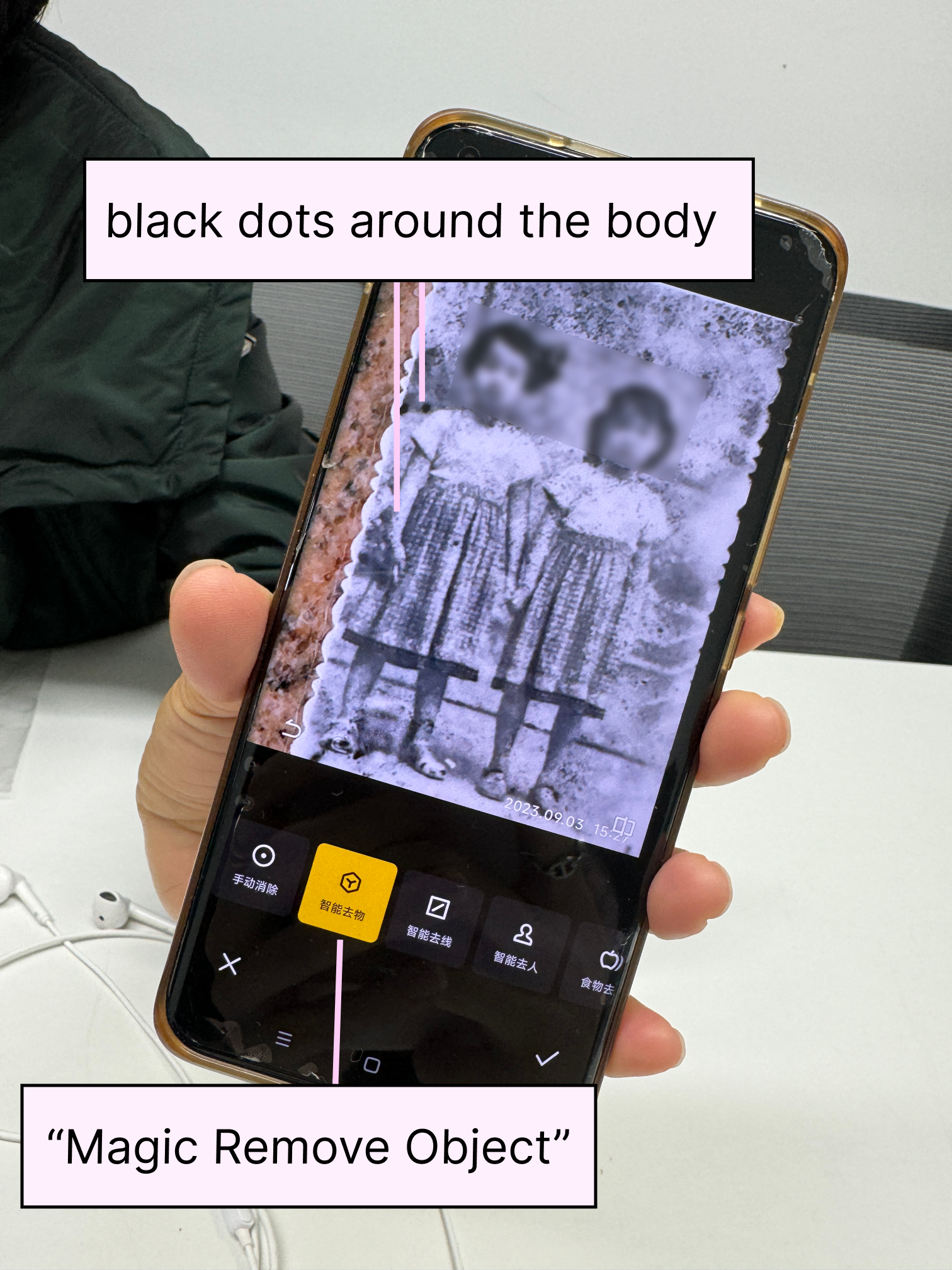}
\caption{P16 demonstrated the photo editing app she uses for manual restoration. Despite the "magic remove object" feature, the multitude of small black dots around the person made the restoration challenging.}
\label{fig:magic-remove-p16}
\end{figure}

However, participants expressed dissatisfaction with existing restoration apps, often due to poor results or high costs. They were willing to invest time in detailed restoration if they had access to suitable and effective tools, which they currently find lacking (see \autoref{fig:magic-remove-p16}). For instance, one participant mentioned, \textit{"I have already repaired those parts... But it's still not working here (around the edge of the body)... it gets distorted, and then I can't do anything about it"} (P16).

P11 emphasized the importance of preserving historical text within photos, seeking tools capable of retaining and restoring these unique details. She mentioned, \textit{"The memorial archway should read ‘first peak in the south of Baiyun Mountain’. It's blurry now, but I wish to retain this information. Preserving textual details, like historical slogans or location names, in photos is ideal."}

\subsubsection{Challenge 2: Long-term Preservation}
A prevalent concern among participants was the long-term preservation of their photos. They used various strategies, like storing digital photos on multiple devices (\textit{"managing"}) or saving photos sent by others in specific social media features (\textit{"collecting"}) that they considered as \textit{"permanent"} (P14). Despite their efforts, they worried about accidental losses, as one participant stated, \textit{"What if I lose my phone or press the wrong button, and everything disappears? I'd like a way to upload to the cloud, maybe?"} (P19). A few other participants shared this concern and viewed cloud storage as a secure and seemingly infinite space for photo preservation. However, many lacked understanding of how cloud storage works and how to use it effectively, leading to uncertainty about the future legacy of their images. For example, P16 pondered the future custodianship of her family's photos, saying, \textit{"I'm not sure if my son will keep them, but I feel it's important to preserve my mother's photos for my generation. Ideally, there would be software that prevents deletion and facilitates their continuous passing down."}

\subsubsection{Challenge 3: Photo Annotation}
\begin{figure}
    \centering
    \includegraphics[width=1\linewidth]{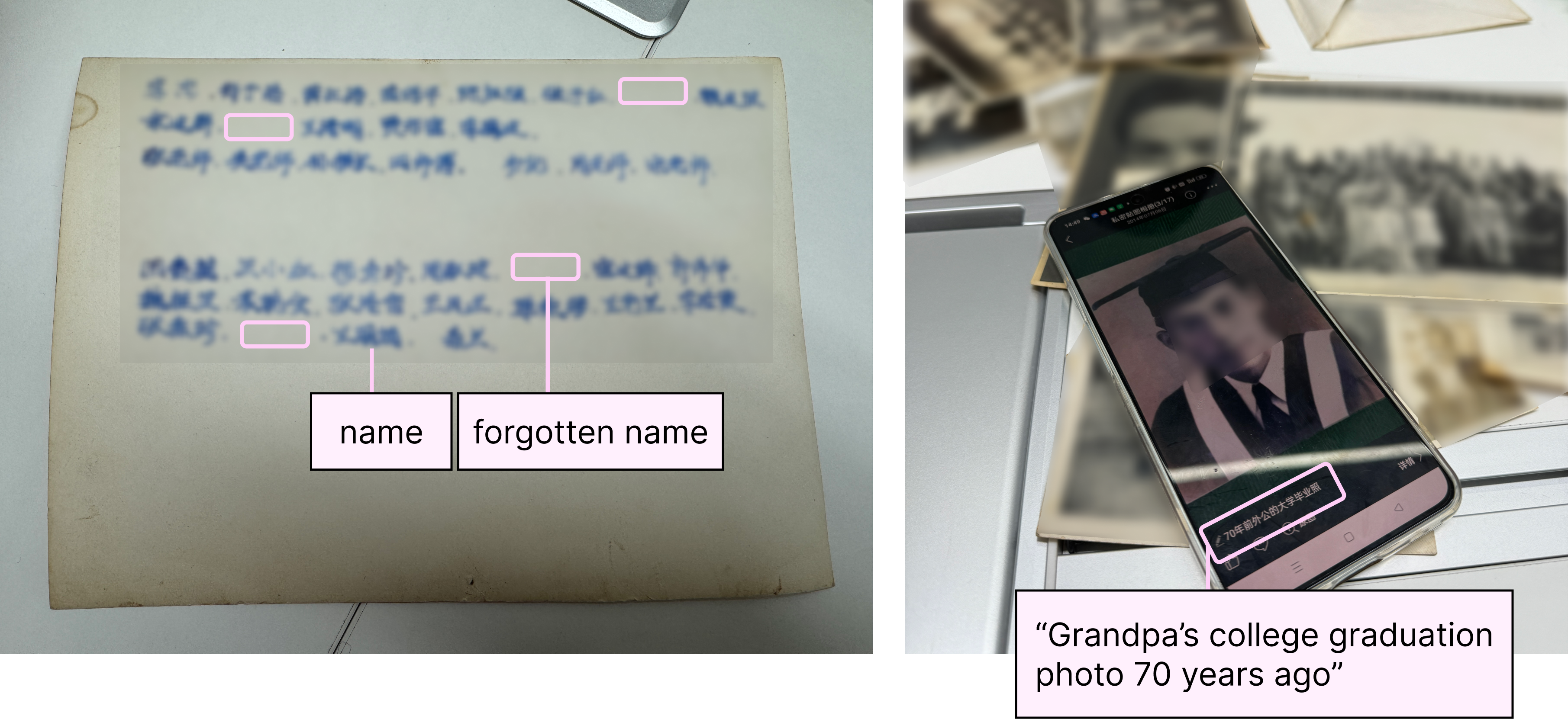}
    \caption{Examples of annotating efforts: P13 labels names on the back of a group photo during memory fading, with highlighted spaces for forgotten names (left); P15 digitized some old photos for storage in the digital photo album within a social media app, adding descriptive text annotations (right).}
    \label{fig:annotation}
\end{figure}
Many participants expressed regret over not documenting details on their photos when they were taken, recognizing its impact on memory recall. This relates specifically to annotation activities \textit{("organizing")}. P1 shared his struggles with remembering specific dates, saying, \textit{"I deeply regret not documenting significant events. Now, I struggle with exact timings, though I generally know the era" }(P1). Similarly, P5 lamented not labeling names on graduation pictures, stating, \textit{"We used to take graduation pictures, and some classmates wanted to write their names on the back. I thought it unnecessary then, but now I can’t recall most of their names… As you age, memory isn't reliable; having name labels would be helpful" }(P5).

P13 started annotating photos upon realizing her fading memory, as shown in \autoref{fig:annotation}, left image. P13 talked about the necessity of this activity, saying, \textit{"It was after many years. Realizing that I don't recognize some people anymore, I hurriedly wrote it down. There are many I can't remember... Gradually, I'm recognizing fewer and fewer people"}.

Manually annotating photos was a common challenge that prevented them from annotating for memory, e.g., \textit{"Back then, we could only use handwritten notes. I stopped annotating in the 1970s, and now, I can't remember the details"} (P11), and \textit{"These group photos, with people clustered together, made it really hard to label their names. So I did not do it... I certainly wanted to but couldn't manage"} (P17).

Even after digitizing their photos, annotating them remained a challenge. P15, for instance, could only annotate a few in the digital photo album within a social media app due to the time-consuming process, as depicted in \autoref{fig:annotation}, right image. P15 described the experience, saying, \textit{"There were some (photos) that I thought were pretty good, and that's how I took them all with my mobile phone and annotated some in Qzone\footnote{Qzone is a social networking application in China, known for its key features like writing blogs and sharing photos.}, but the process was too time-consuming to do for all my photos"}.

\subsubsection{Challenge 4: Photo Retrieval}
\begin{figure}
\centering
\includegraphics[width=0.7\linewidth]{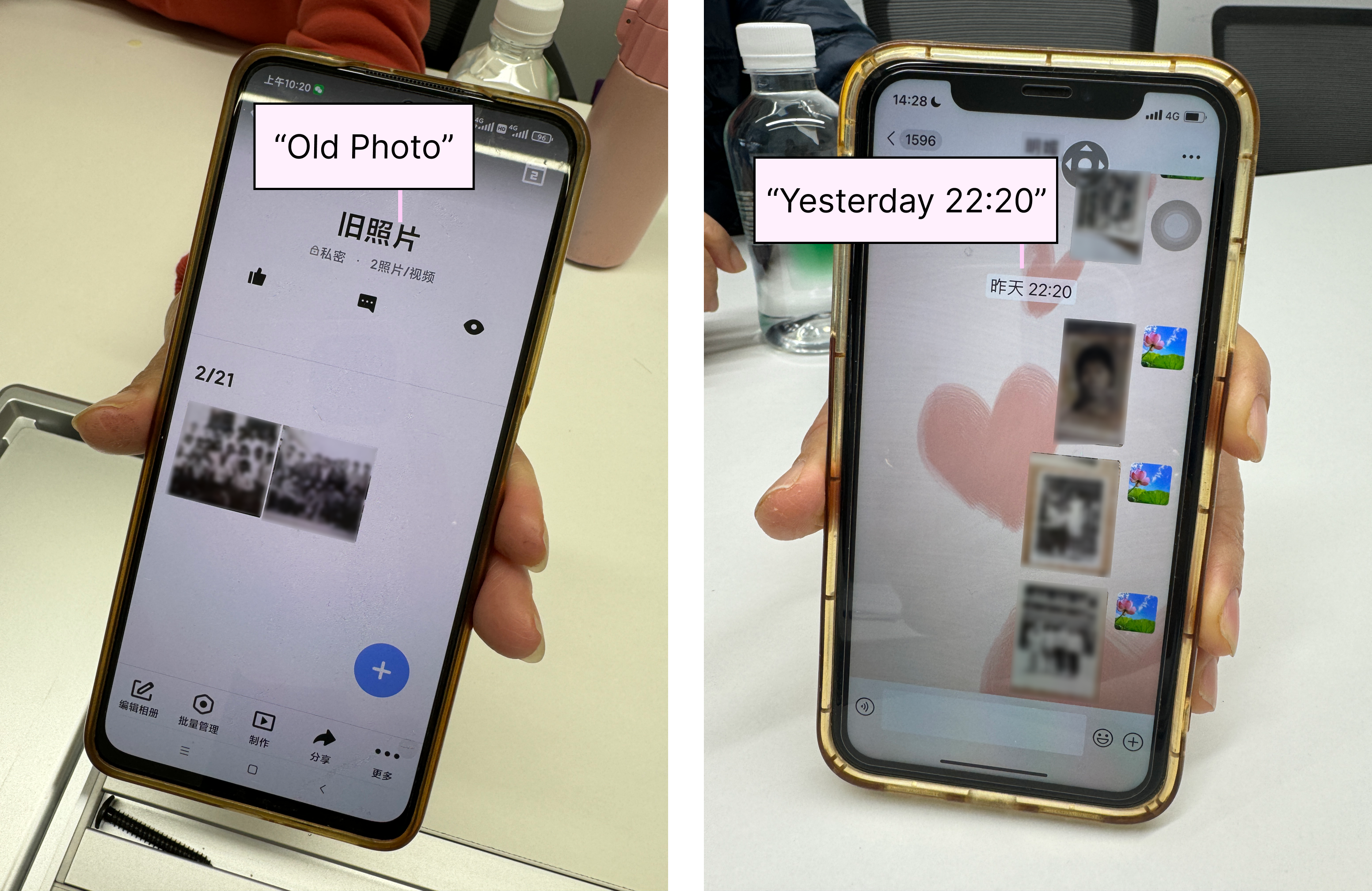}
\caption{Examples of digital sharing strategies for a selected set of photos: P19's private digital album (left) and P20's private WeChat messages (right). Each strategy comes with its own set of challenges: P19 struggles with efficiently organizing her digital album, while P20's practice of daily photo deletion—although not depicted here—stems from frustration with scrolling through thousands of photos. This routine helps P20 maintain a fresh and manageable collection for reminiscence but also requires continuous curation of new content.}
\label{fig:two-digital-sharing}
\end{figure}
Participants reported difficulties in locating specific photos within their phone albums or group chats, especially during the process of sharing and storytelling. Participants P19 and P20 demonstrated different strategies for creating accessible sharing spaces (\textit{"triaging"}).

P19 created a private album for her photos but faced challenges in compiling all the desired images due to the cumbersome retrieval process. As P19 mentioned, \textit{"I uploaded some photos last night to this album, but it was too cumbersome to find and include all the ones I wanted"}.

P20, less proficient in managing photos, adopted a simpler method by sending images to herself in a private WeChat\footnote{WeChat is a social media application in China with its main feature being instant messaging.} message. This approach, prevalent among our participants, offered a temporary solution for immediate viewing. As explained by P20, \textit{"I selected these from the album, but then didn’t know where else to put them"}. However, P20 also mentioned the inconvenience of navigating through a vast number of photos in this private space (\textit{"browsing"}), often leading to their deletion after viewing. P20 explained, \textit{"Scrolling through thousands of photos is frustrating. I post them in my WeChat for a brief look, then delete them"}.

\subsection{Participatory Design Results for Photo-based Reminiscence}
\begin{figure}
    \centering
    \includegraphics[width=0.75\linewidth]{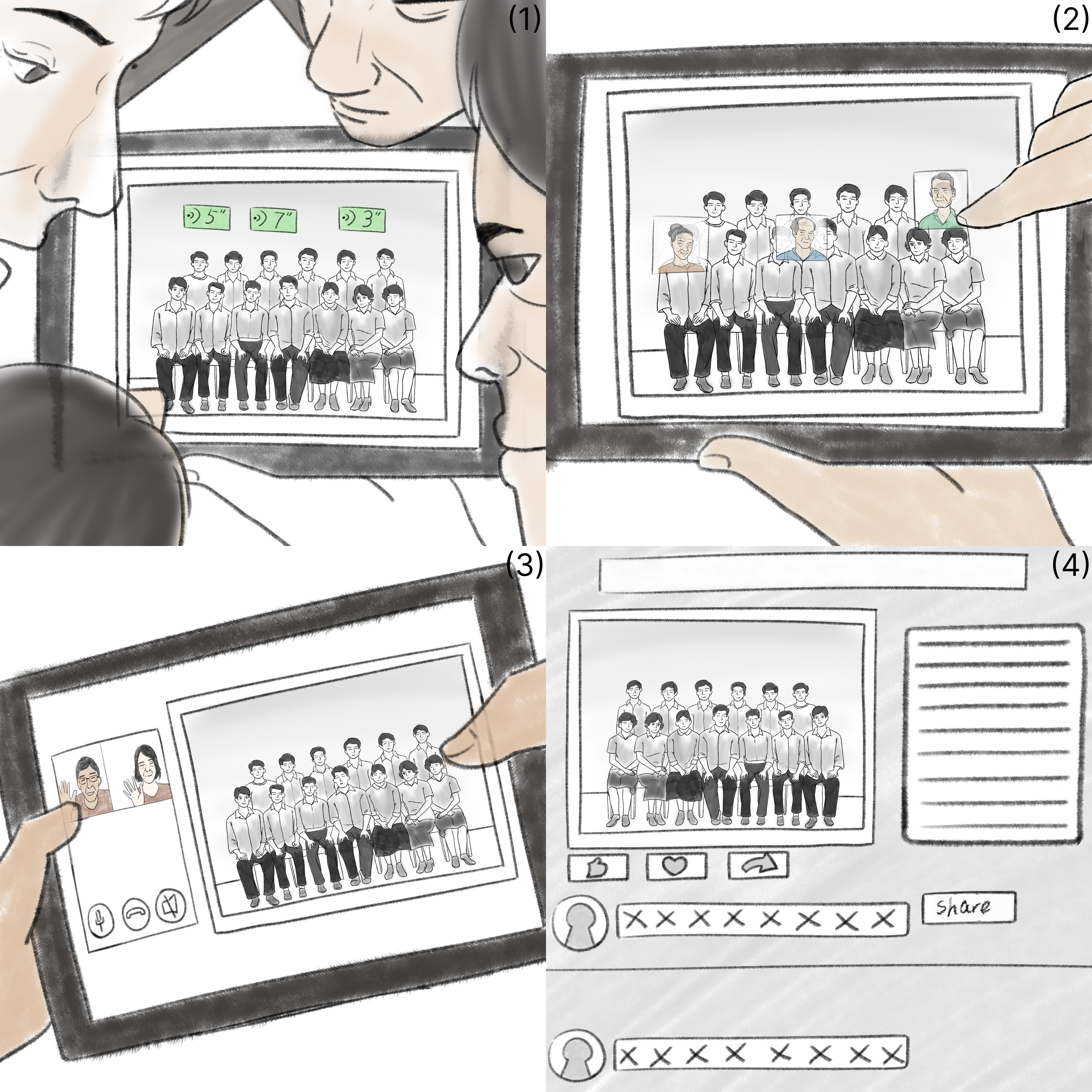}
    \caption{Older adults' co-design ideas on collective reminiscing with old group photos, illustrated by the co-author. (1) \textbf{Co-creating an Artifact} - This concept envisions family or friends gathering to add voice recordings to images, thereby annotating and capturing their memories associated with the picture. (2) \textbf{Asynchronous Familiarization and Recognition} - This idea involves combining past and present images of a person and providing an option to listen to their voice recordings. This feature aims to help refresh memories of old acquaintances by bridging past and present connections. (3) \textbf{Integrating Direct Contact Options in Photos} - This concept transforms old photos into a video call directory. By clicking on a photo, users can initiate a call with old friends, seamlessly connecting memories with real-time interaction. (4) \textbf{Reconnecting with Old Acquaintances} - This idea allows users to upload photos and information to search and reunite with old acquaintances they've lost contact with, facilitating the sharing of memories and rekindling of past relationships.}
    \label{fig:collab}
\end{figure}
From the participatory design activities, we identified four major subthemes that older adults desire in digital tools supporting photo-based reminiscence, encompassing aspects of photo sharing, viewing, and annotating. Specifically, these include: 1) Collective Reminiscing with Group Photos, 2) Photo as a Legacy for Future Generations, 3) Flexible Display for Browsing, Viewing, and Sharing of Photos, and 4) Modalities in Photo Annotation.

\subsubsection{Collective Reminiscing with Group Photos}
Participants P7, P8, P13, P14, and P20, who brought group photos from their past, such as old classmates and colleagues, inspired ideas for group reminiscence (see \autoref{fig:collab}). They emphasized the importance of collaborative features, including attaching comments and memories to photos with different motivations.

\textbf{Co-creating an Artifact.} P7 suggested each individual in a group photo could attach a voice recording, akin to a \textit{"comment section"}. This method would allow participants to contribute their personal memories or thoughts to each photo, thereby creating a rich and collective photo artifact:
\begin{quote}
    \textit{"After this image is displayed, one could introduce it (via voice recording) … after the introduction, there could leave a space for others... In that blank space, the person involved can record what they want to say."} (P7)
\end{quote}
Furthermore, P7 emphasized the lasting impact of such a feature:
\begin{quote}
   \textit{ "Your photos should be preserved forever. The time and place of the photo are already known. Then, when you see this photo, whatever thoughts or events you had at that moment, or any interesting occurrences, can be recorded."} (P7)
\end{quote}

\textbf{Asynchronous Familiarization and Recognition.}  
P7, P8, P20 also envisioned using group photos as a means for better recognition in chat applications and in-person scenarios. P7 proposed that old classmates update their appearances on the digital group photo with current selfies to create a composite image that blends the past with the present in the group chat. P8 suggested overlaying current images onto the original for a direct comparison. P20 focused on the practicality of labeling and updating images for easier recognition, especially during reunions:
 \begin{quote}
     \textit{ "Some people you might not recognize their names. But if you could highlight them (in the photo) and see them speaking, then you'll recognize them."} (P20)
 \end{quote}
 P8 proposed a virtual roll call feature drawing metaphors from school experience, adding an interactive layer to the photo-viewing experience: 
 \begin{quote}
      \textit{"You could directly call \lq Wang Ming\rq, and then the person would appear said \lq here!\rq and start talking... It becomes more interesting"}.
 \end{quote}

\textbf{Integrating Direct Contact Options in Photos.} P14 introduced the idea of using group photos as a directory for initiating direct contact, such as voice or video calls. This feature would enable users to click on a person in the photo to start a call:
 \begin{quote}
     \textit{"It would be even better to add an option for voice or video calls. Being able to choose either voice or video calls by clicking on someone in the image would be ideal."} (P14)
 \end{quote}

\textbf{Reconnecting with Old Acquaintances.} P13, interested in reconnecting with old classmates, suggested uploading photos to a platform with additional information like school and year. This feature would facilitate users in finding and reconnecting with each other, sharing comments, and catching up:
\begin{quote}
    \textit{"I want to find (them), and maybe (they) also want to find me. Then, everyone can upload this information, creating a search that can facilitate this connection."} (P13)
\end{quote}

\subsubsection{Photo as a Legacy for Future Generations}
Our study found a strong inclination among participants to make their photos accessible and meaningful to descendants, aiming to leave reminiscing materials for future generations.
 
\textbf{Adding Personally Relevant Historical Context.}  Participants expressed a preference for adding content to photos that is not only relevant for inter-generational understanding but also personally significant to them. A significant number of participants (N=10) valued the addition of historical context to photos, believing it not only enriches their memories but also provides insightful content for future generations: 
 \begin{quote}
      \textit{"1989, a time of turmoil... it's meaningful to juxtapose our wedding photo with an image of a major event from that year."} (P8) 
 \end{quote}
 Augmenting images with historical contexts is valuable, yet it's crucial to ensure that these additions are personally significant to older adults:
 \begin{quote}
\textit{"I can search for historical event photos myself. They aren't necessary in family albums if I didn't personally experience them."} (P6)
\end{quote}
 
 When the context is not visually related to the image, they favor text annotations over additional photographic elements:
\begin{quote}
\textit{"Aligning my birthday with significant global events, like China joining the United Nations, adds depth to my personal history...in this case, adding only text is enough."} (P17)
\end{quote}
 
\textbf{Constructing a Family Narrative with Multi-media Format.} P18 proposed a tool with editable templates for documenting family history, values, and traditions through photos. This tool would allow for creating short videos or slideshows that encapsulate family stories, appealing to younger generations and preserving family legacies:
 \begin{quote}
     \textit{"You could develop a editable template based on this concept, which could then be applied to every family. Each family would include grandparents, then parents, and then the current generation, including grandchildren. It would document things like how they were raised from childhood, their virtues, and how these are passed down...The theme of family traditions and how they are passed down could be captured in a short video or a similar format.  These could be turned into videos or slideshows, enhanced with music, to pass on meaningful family narratives...Music can change based on the ups and downs of the story."} (P18)
 \end{quote}
 P9 also expressed ideas for creating a family narrative by suggesting a series of photos showcasing family changes over the years, accompanied by narratives and fitting music:
\begin{quote}
    \textit{"If we are living comfortably or are relatively affluent, our expressions are like this. In years when we are struggling and facing hardships, our expressions would change...all can be reflected...Creating something for our descendants to see the changes we've gone through over the years would be valuable. Looking at our expressions in photos from each year, I can also recall our living conditions at the time, right?... alongside a narrative and music, would be a precious legacy."} (P9)
\end{quote}
P17 proposed including objects that represent family history, such as family letters, in photos to add depth to the narrative.

\subsubsection{Flexible Photo Display for Browsing, Viewing, and Sharing of Photos}
 While acknowledging the convenience of digital albums on mobile phones (\autoref{fig:cards-overview} (A)), many expressed a stronger inclination towards more traditional, nostalgic formats for digital reminiscing. A notable preference among participants was for long film album design (\autoref{fig:cards-overview} (C)) and traditional photo albums design (\autoref{fig:cards-overview} (D)).

\textbf{Individual Photo Display.} The preference for the long film album designs is primarily due to their ability to display photos in a focused, individual manner:
\begin{quote}
    \textit{"If saving digitally, there’s no need for cluttered interfaces. It’s better to view each photo individually."} (P6)
\end{quote}
\begin{quote}
    \textit{"Photos should be viewed one at a time. When I want to reminisce about a specific photo, I should be able to focus on just that one, without the confusion of multiple images."} (P13)
\end{quote}
This preference also extends to those favoring traditional photo albums designs, emphasizing the desire for an option to zoom in on specific images for a more intimate reminiscing experience.

\textbf{Cinematic Experience with Auto-play Feature.}
Several participants proposed an auto-play feature for a cinematic experience, where photos flow horizontally like a film roll that \textit{"portrays the feeling of the passage of time"} (P17), with the option to pause on a specific image:
\begin{quote}
    \textit{"Can it operate automatically like a film roll? Once it's automatic, I see it flowing. If I want to see a specific photo, I can just click on it to stop the flow, which would be nice."} (P1)
\end{quote}

P11 and P18 suggested developing templates for authoring videos, especially useful when photos are part of a series.

\textbf{Contextual Display.}
Participants expressed a strong appreciation for the nostalgic and visual contextual value inherent in traditional album designs. They indicated that this design enables additional clues within these albums to assist in deducing information, especially helpful when encountering memory loss. This reflects a need for more subtle yet supportive elements in memory aids that cater to the challenges of memory recall:
\begin{quote}
    \textit{"By seeing people of the same \lq type\rq, I can know that, for example, they are from a certain company. Even if I get forgetful, I would still recognize that they belong to a specific organization. I wouldn't have to struggle to remember which factory or company they worked at, as they wouldn't all be mixed together."} (P16)
\end{quote}
P20 highlighted the importance of tailoring photo groupings to different scenarios and individuals for more meaningful sharing:
\begin{quote}
    \textit{"Different scenarios, like reunions or family gatherings, should have corresponding photo groupings, so ensure that everyone feels the relevance."} (P20)
\end{quote}

\textbf{Preserving Physical Context and Properties.} The physical connections of photos to objects and documents previously attached to were also emphasized, with a preference for digital album formats that reflect original photo sizes and contexts:
\begin{quote}
    \textit{"Like the one-inch small photos we had back then, you could tell they were for work or identification purposes. But if all the photos are made the same size like now, it's unclear what the purpose of each photo is."} (P16)
\end{quote}

\subsubsection{Modalities in Photo Annotation}

In the realm of photo annotation, our study sheds light on the significance of various factors influencing participants' preferences, emphasizing not only the ease of recording memories but also the quality of future revisits, both personally and by others. While participants appreciate the intuitiveness of voice-to-text features, concerns about accuracy persist, as highlighted by issues with pronunciation (P1) and challenges in correcting errors (P9), leading to a preference for combining voice input and manual text editing (P13). Voice input is valued for its expressiveness, adding depth to the photo experience (P1), though participants acknowledge limitations in quiet environments (P9) and stress the importance of atmosphere for genuine emotional expression (P17). Emotional connections vary, with some preferring handwriting for photos with profound significance, despite its labor-intensive nature, for its personal touch and nostalgic evocation. However, handwriting is noted for its inefficiency and accessibility issues. Concerns about the longevity of voice annotations compared to text reflect a desire for permanent preservation of photos and their annotations for future generations, as expressed by P20. Inclusiveness is crucial, with participants advocating for multiple annotation options to accommodate different abilities and preferences, such as typing for those who find speaking challenging (P5) and voice input for those with deteriorating eyesight (P11), ensuring effective photo annotation for users with diverse needs.
 
\section{Discussion and Implications}
In this section, drawing from semi-structured interviews on older adults’ daily photo-based reminiscence practices, their experiences and challenges with existing digital tools, and co-design activities envisioning their ideal digital reminiscence experience, we outline design implications from our findings to inform the development of reminiscence technologies for older adults. These include improving accessibility in digital photo tools, balancing memory prompts with emotional well-being, fostering social connectivity through interactive photo sharing, supporting creative expression with intuitive storytelling tools, and integrating personal and cultural-historical contexts to enrich reminiscence. Together, these insights provide a comprehensive foundation for designing reminiscence technologies that address the specific needs of older adults.

\subsection{Towards Accessible Reminiscence for Older Adults: Improving Digital Photo Tools to Support Realistic Reminiscence Activities}
Digital technology designed to help older adults reminisce often employs photos and verbal prompts to evoke memories for exercises or social interactions. Our investigation highlights several unmet needs in how older adults interact with these cues in daily life. These needs can hinder their engagement in reminiscence activities, which are crucial for memory exercise and social sharing. 

Our findings (Sections 4.1.1 and 4.3.4) highlight a significant challenge: the potential loss of memory cues, including physical photos, digital copies, and annotations. This loss can profoundly impact older adults' sense of self \cite{lindley_placing_2015, sas_exploring_2018}. While older adults employ various technologies and strategies to digitize their photos for preservation, concerns about losing these memories in unfamiliar systems persist (Sections 4.1.2 and 4.2.3). Additionally, deteriorating image quality can result in the loss of meaningful photo details (Section 4.2.2). Creating applications that aid in preserving various memory cues—from different photo formats to image quality and annotations—could potentially enhance older adults' ability to reminisce. Given that older adults have different mental models for photo tools compared to younger adults \cite{axtell_back_2019}, it is recommended that designers understand and accommodate these differences. By carefully considering older adults' mental models with different photo interactions, designers can create more usable and effective tools for reminiscence.

Older adults often struggle to curate and organize images scattered across various digital platforms, making it difficult to retrieve specific memories (Sections 4.1.2, 4.2.4, and 4.2.5). Despite strategies like creating digital albums or sending private messages to themselves, effective photo management remains challenging. Digital photo annotation is crucial for memory support and efficient search capabilities \cite{ames_why_2007}. Although early perspectives suggested limited need for extensive annotation due to context and recency \cite{rodden_how_2003}, our research highlights its growing importance for older adults as the temporal distance from their photo memories increases. Older adults consider annotating time, location, people, and events critical, yet they often find current manual annotation tools cumbersome. Existing systems based on timestamps \cite{graham_time_2002, chen_exploring_2023} or location data \cite{mcgookin_reveal_2019} may not be adequate for older photos that lack digital metadata.

Integrating the need for photo retrieval and annotation, and considering older adults' storytelling patterns in photo activities (Section 4.1.5), we suggest leveraging their storytelling propensity to enhance photo management processes. Inspired by PhotoFlow \cite{axtell_photoflow_2019}, which clusters photos using speech from oral storytelling, we propose using narrative reminiscence to automatically annotate photos. This approach would involve older adults' deductive process of piecing together contextual clues from photos, recording details such as time, location, people, and events, thereby aiding retrieval and preserving oral histories. Narratives, both in voice and text, could be archived for future visits. To ensure accuracy, these narratives should allow for editing and provide multimodal annotation options, considering the factors presented in Section 4.3.4.

Interactive and engaging annotation methods can further support this process. Prior studies converting photo annotation into interactive activities via online chats \cite{qian_exploring_2004} suggest the potential of conversational agents. These agents, paired with advanced language models, could streamline the annotation process and engage older adults effectively \cite{ding_talktive_2022, simpson_daisy_2020}. Additionally, robots with dialogue systems designed for photo storytelling \cite{tokunaga_dialogue-based_2021} present promising possibilities. These methods could simplify retrieval and transform photos into cohesive narratives, enhancing photo management experiences for older adults.

\subsection{Towards Sensitive Reminiscence for Older Adults: Balancing Memory Prompts with Emotional Well-being}
Prior quantitative research suggests that digital tools for viewing photos—such as galleries, slideshows, and tabletops—vary in their effectiveness at prompting memories \cite{axtell_design_2022}. Gallery views, displaying many small, cropped images, are less effective for reminiscence due to the lack of visible details. Conversely, slideshows presenting one clear, large image at a time serve as stronger memory prompts but often evoke factual rather than personal memories. Tabletops, which provide full details and context of each image, offer a balanced approach by promoting more personal memories without overly increasing factual ones.

Our findings (Sections 4.1.5 and 4.3.3) support the benefits of tabletop views, indicating that participants prefer having sufficient clues to independently reconstruct forgotten details. This aligns with previous work showing that older adults with cognitive impairment prefer receiving just the right amount of cuing assistance from caregivers, enabling them to recall most of the memory themselves \cite{lee_providing_2007}. In our study, digital tools that provide independent cuing assistance by grouping images by theme—such as colleagues from the same company—or by original size and associated physical documents can suggest their context and purpose, offering a more sensitive way to help older adults recall memories. Contextual display designs like tabletop displays can facilitate this approach. Given that slideshows and gallery views remain predominant in digital picture interaction \cite{axtell_design_2022}, we recommend exploring and implementing contextual displays like thematic clusters (e.g., family or familiar places) and tabletop views to provide contextually rich memory prompts that better support reminiscence among older adults.

In addition to providing effective memory prompts, it is equally critical to address the emotional dimensions of reminiscence. As detailed in Section 4.1.4, reminiscence can elicit both positive and negative emotions, as noted in prior research \cite{mongrain_positive_2012, cappeliez_empirical_2006}.In addition to positive reflections, we observed instances of sadness tied to negative reminiscence, such as Intimacy Maintenance \cite{cappeliez_empirical_2006}, reflecting on the loss of a loved one, and Bitterness Revival \cite{cappeliez_empirical_2006}, revisiting challenging life experiences. Such emotions, if unresolved, may disrupt well-being and even sleep, given that reminiscence often arises spontaneously and without evaluative framing \cite{staudinger_life_2001}, which can prolong the emotional impact.

Consequently, designing reminiscence technologies for older adults should encompass not only the goal of eliciting richer or more sustained memories but also address the emotional nuances these interactions may invoke. For instance, providing options to organize photos by emotional significance, rather than solely chronological or thematic order, could enable users to regulate the emotional intensity of their experiences, emphasizing images that offer comfort while managing access to potentially distressing ones. Future designs could also incorporate personalized emotion-detection features, allowing systems to adjust prompts based on real-time emotional cues, such as facial expressions or vocal tones. In situations where negative emotions arise, or emotions become overwhelming, features that modulate prompt intensity—such as “lighter” or “deeper” reminiscence modes—could be beneficial. For example, slideshows, which evoke more factual memories, may be well-suited for less intensive reminiscence experiences. Additionally, as discussed in Section 4.1.4, older adults often transform negative emotions into positive reflections through “redemption sequences,” where challenging memories are reframed constructively \cite{mcadams_when_2001}. Designing reminiscence activities that support this emotional processing—such as life story creation or guided memory reviews to foster redemption sequences—can promote emotional well-being \cite{wildschut_nostalgia_2006}. With the recent increase in conversational agents facilitating reminiscence \cite{jin_exploring_2024}, further research might explore context-sensitive AI-driven interactions, allowing deeper yet moderated memory exploration.  We recommend further exploration in this area to empower older adults in managing the emotional impact of reminiscence, fostering a balanced and sensitive experience.

\subsection{Towards Connected Reminiscence for Older Adults: Enhancing Social Connectivity through Photo Sharing}
Enhancing social connectivity through photo sharing is a priority for participants, who seek to overcome geographical barriers by using photos to revive shared memories and engage in collaborative activities that bridge physical distances (Section 4.3.1). Photos are valued not only for direct interaction but also as tools for remote collective experiences, creating multimedia artifacts that preserve memories and foster social interaction. We recommend adding more interactive and socially driven features to photo tools to meet the social needs of older adults. For example, Ito et al. developed "PhotoChat," a system designed for real-time, collaborative interactions through annotated photos with handwritten details \cite{ito_personal_2006}. This system demonstrates the potential of digital photos to serve as conversation starters and memory triggers in social settings, fostering deeper connections through shared visual experiences. Although handwritten annotations in PhotoChat might not be accessible for all older adults, participants suggested attaching voice recordings to images as an alternative. This approach would allow different people in the image to share their memories and communicate through it, providing a more inclusive option. Group photos, which often invoke collective reminiscence but are typically limited in ownership as noted in Section 4.1.3, could particularly benefit from this remote and collective approach.

Additionally, participants expressed a keen interest in immersive, nostalgia-driven environments. We propose technologies that could enhance the immersiveness of the experience. VR offers a unique platform for simulating collocated photo-sharing activities in a virtual space \cite{li_measuring_2019}, similar to the nostalgic experiences enjoyed by older adults in projects like School Days \cite{baker_schools_2021}. Besides reminiscing with strangers in School Days, we also encourage exploring the potential for facilitating remote interactions among friends or acquaintances, with whom older adults reminisce most frequently according to our findings (Section 4.1.1) and prior studies \cite{bluck_reminiscence_1998}. Additionally, different social functions of reminiscence should be considered, as reminiscing for identity is more common with familiar individuals like partners, while reminiscing with strangers and children often focuses on teaching and informing \cite{bluck_reminiscence_1998}. Recognizing these distinct social functions can help tailor reminiscence technologies to better support the emotional and social needs of older adults.

\subsection{Towards Creative Reminiscence for Older Adults: Enabling Memory Authoring and Storytelling}
The emergence of new tools with low barriers to entry has satisfied the needs of older adults for memory authoring and creative expression. Our findings (Section 4.1.3) reveal a new phenomenon: the ease of using editing software on short video sharing platforms (SVSPs) has led older adults to create short videos for reminiscence. In the memory video-making process, older adults curate a series of photos, carefully select music that matches the atmosphere and stories, and use AI-generated tags as inspiration to write captions for sharing. They derive significant enjoyment from creating a narrative experience that is pleasing to view and share.

This form of sharing is also valued as an experience that can be passed down to future generations, as mentioned in Section 4.3.2. This finding aligns with prior research indicating that older adults hope to leave reminiscing materials for future generations \cite{lindley_before_2012, jones_co-constructing_2018-1}. Participants aim not only to preserve photos but also to convey the stories of different family members and family traditions through an integrated narrative experience. 

This observation aligns with previous work \cite{tang_towards_2023}, which indicated that the low barrier to video creation via SVSPs empowers Chinese older adults to engage actively in video creation. However, our findings differ in that, unlike the high involvement of Chinese older adults in posting videos on SVSPs, our participants exhibited a more conservative approach to sharing videos made from family photos, primarily due to privacy concerns.

Our findings emphasize that memory technologies should support creativity and meaning-making rather than passively providing memory cues \cite{petrelli_making_2009}. Thiry et al. designed a memory authoring tool that allows older adults to build personal timelines with photos, achieving meaning through active selection and creativity \cite{thiry_authoring_2013}. Follow-up work further proposed the concept of visual mementos for general users, using complex visualizations of personal data for reminiscence and sharing of life experiences \cite{thudt_visual_2016}. While current reminiscence tools for older adults primarily focus on cognitive engagement and social interaction (e.g., \cite{tokunaga_dialogue-based_2021, baker_schools_2021, li_exploring_2023}), tools that empower older adults in memory authoring remain relatively under-explored. Expanding these tools to support personal meaning-making can empower older adults as active authors of their own narratives, fostering opportunities for content creation and self-expression and aligning with broader goals of late-life development \cite{brewer_tell_2016, rogers_never_2014}.

We suggest developing memory authoring tools that support older adults in creating and sharing content such as series of photos in video formats, prioritizing usability and accessibility for those with varying technical skills. Further work can provide a more detailed overview of the challenges older adults face in this process. Essential features could include easy photo selection, narrative optimization, and accessible templates for reminiscing \cite{thiry_authoring_2013}. Additionally, tools for choosing music, enhancing image quality, and adding contextual details can enrich their stories. For stories intended for future generations, the focus should be on educational content that conveys family spirit through pictures, music, and text. Tools that help older adults derive themes and sentiments from photos can enhance storytelling by moving beyond geographical or temporal information to include aspects like sentiment and expressions. Another important aspect is supporting older adults in tailoring their videos for different audiences, balancing the tension between authenticity and educational value \cite{thiry_authoring_2013}. Flexible tools that cater to various audiences can be further explored.

\subsection{Towards Culturally-Aware Reminiscence for Older Adults: Integrating Personal and Historical Contexts} 
Our findings from Sections 4.1.5 and 4.3.1 indicate that older adults have specific content preferences and methods for recalling memories, as well as a preference for collective reminiscence activities supported by technology. It is essential to connect these characteristics and preferences to a broader social context. Previous research in memory studies suggests that collective reminiscence, as a socially shareable memory system, must consider the characteristics of the community and cultural-historical context, which significantly influence how individuals and communities remember and value the past \cite{wang_cultural_2008}. Cultural differences shape the processes, practices, and outcomes of collective memory, potentially impacting the effectiveness and acceptance of memory technologies across different cultural contexts. However, this cultural perspective on memory is often overlooked in current HCI research.

One important aspect is the variation in how people from different cultures integrate personal memories with historical events. In a study by Wang and Conway \cite{wang_stories_2004}, approximately 17\% of the autobiographical memories recalled by Chinese participants were situated within a historical context, compared to less than 2\% of the memories from European American participants. This difference highlights how Chinese culture emphasizes collective history and shared experiences, influencing individuals to connect their personal memories with broader historical narratives. Reflecting this in our study (Section 4.3.2), participants preferred organizing their photos within a historical context, such as aligning personal photos with major global or national events from the same year, to cherish their personal history.

Another important consideration is the variation in how people from different cultures incorporate collective identity into individual self-conception. Empirical studies \cite{cousins_culture_1989, trafimow_tests_1991, wang_culture_2001} have shown that European and European American adults tend to focus on their unique personal attributes and qualities, whereas Asians often refer to their social roles, important relationships, and group memberships. Our findings (Section 4.3.1) reflect that participants enjoyed creating social photo experience where everyone could contribute to constructing an artifact for collective memory and maintaining social relationships.

We recommend that future HCI and CSCW researchers and designers consider the social and cultural contexts when designing memory technologies for older adults. This cultural distinction suggests that memory technologies for Chinese users could facilitate the integration of personal and historical events, allowing users to contextualize their memories within significant cultural and historical frameworks. These technologies should support the documentation and sharing of memories that emphasize social connections and collective experiences. Conversely, memory technologies designed for Western users might focus on enhancing individual self-expression and personal storytelling. Providing tools that allow users to curate and share their unique life stories, personal reflections, and individual perspectives can support this goal. We encourage future work to continue exploring how these variations can impact the design of memory technologies.

\section{Limitations}

This study was conducted within an urban community in China, involving participants with diverse backgrounds, origins, and pre-retirement occupations. While this diversity was considered, it is essential to note that our findings predominantly reflect the cultural nuances of an oriental context. This limitation may affect the generalizability of our results to a broader population of older adults with varying cultural and socioeconomic backgrounds. Future research should aim to include a more diverse sample to better understand the needs and preferences of a wider range of older adults. Additionally, the activities older adults engage in are influenced by the functionalities available in digital tools. Future studies should analyze the activities that older adults from different regions engage in and the functionalities accessible to them in these digital tools. This will provide a more comprehensive understanding of how digital tools can be designed to support reminiscence activities across various cultural and regional contexts. Lastly, while our study offers significant insights into the photo activities older adults engage in and the general challenges they face, we acknowledge that our approach may not be exhaustive. Older adults might not recall every activity they engage in, leading to potential oversights. We recommend that future research further investigate these aspects, potentially employing more rigorous methods such as having participants demonstrate their activities in simulated settings or conducting longer-term studies. These approaches would help achieve more comprehensive coverage and provide deeper insights into their photo-based reminiscence practices.

\section{Conclusion}

We present a qualitative study involving semi-structured interviews and co-design activities with 20 older adults in China. Our findings reveal the ways these individuals reminisce with photos in daily life and the challenges they encounter when using digital tools for photo-based reminiscence. Digital tools provide valuable opportunities for photo self-viewing and sharing, enhancing the reminiscence experiences of older adults. Based on the findings, we recommend designing reminiscence technologies that prioritize preservation and accessibility, incorporate sensitive and emotionally balanced interactions, enhance social connectivity, support creative expression, and consider cultural contexts to better support the photo-based reminiscence activities of older adults.


\bibliographystyle{ACM-Reference-Format}
\bibliography{references}


\begin{thebibliography}{72}

%%% the Web address.

\ifx \showCODEN    \undefined \def \showCODEN     #1{\unskip}     \fi
\ifx \showDOI      \undefined \def \showDOI       #1{#1}\fi
\ifx \showISBNx    \undefined \def \showISBNx     #1{\unskip}     \fi
\ifx \showISBNxiii \undefined \def \showISBNxiii  #1{\unskip}     \fi
\ifx \showISSN     \undefined \def \showISSN      #1{\unskip}     \fi
\ifx \showLCCN     \undefined \def \showLCCN      #1{\unskip}     \fi
\ifx \shownote     \undefined \def \shownote      #1{#1}          \fi
\ifx \showarticletitle \undefined \def \showarticletitle #1{#1}   \fi
\ifx \showURL      \undefined \def \showURL       {\relax}        \fi
% invisible to TeX
\providecommand\bibfield[2]{#2}
\providecommand\bibinfo[2]{#2}
\providecommand\natexlab[1]{#1}
\providecommand\showeprint[2][]{arXiv:#2}

\bibitem[Ames and Naaman(2007)]%
        {ames_why_2007}
\bibfield{author}{\bibinfo{person}{Morgan Ames} {and} \bibinfo{person}{Mor Naaman}.} \bibinfo{year}{2007}\natexlab{}.
\newblock \showarticletitle{Why we tag: motivations for annotation in mobile and online media}. In \bibinfo{booktitle}{\emph{Proceedings of the {SIGCHI} {Conference} on {Human} {Factors} in {Computing} {Systems}}}. \bibinfo{publisher}{ACM}, \bibinfo{address}{San Jose California USA}, \bibinfo{pages}{971--980}.
\newblock
\showISBNx{978-1-59593-593-9}
\urldef\tempurl%
\url{https://doi.org/10.1145/1240624.1240772}
\showDOI{\tempurl}


\bibitem[Axtell et~al\mbox{.}(2023)]%
        {axtell_underdeveloped_2023}
\bibfield{author}{\bibinfo{person}{Benett Axtell}, \bibinfo{person}{Eleen Gong}, {and} \bibinfo{person}{Cosmin Munteanu}.} \bibinfo{year}{2023}\natexlab{}.
\newblock \showarticletitle{An {Underdeveloped} {Metaphor}: {The} {Mismatched} {Designs} and {Motivations} of {Digital} {Picture} {Interactions}}.
\newblock \bibinfo{journal}{\emph{ACM Transactions on Computer-Human Interaction}} \bibinfo{volume}{30}, \bibinfo{number}{2} (\bibinfo{date}{April} \bibinfo{year}{2023}), \bibinfo{pages}{1--36}.
\newblock
\showISSN{1073-0516, 1557-7325}
\urldef\tempurl%
\url{https://doi.org/10.1145/3569887}
\showDOI{\tempurl}


\bibitem[Axtell and Munteanu(2019a)]%
        {axtell_back_2019}
\bibfield{author}{\bibinfo{person}{Benett Axtell} {and} \bibinfo{person}{Cosmin Munteanu}.} \bibinfo{year}{2019}\natexlab{a}.
\newblock \showarticletitle{Back to {Real} {Pictures}: {A} {Cross}-generational {Understanding} of {Users}' {Mental} {Models} of {Photo} {Cloud} {Storage}}.
\newblock \bibinfo{journal}{\emph{Proceedings of the ACM on Interactive, Mobile, Wearable and Ubiquitous Technologies}} \bibinfo{volume}{3}, \bibinfo{number}{3} (\bibinfo{date}{Sept.} \bibinfo{year}{2019}), \bibinfo{pages}{1--24}.
\newblock
\showISSN{2474-9567}
\urldef\tempurl%
\url{https://doi.org/10.1145/3351232}
\showDOI{\tempurl}


\bibitem[Axtell and Munteanu(2019b)]%
        {axtell_photoflow_2019}
\bibfield{author}{\bibinfo{person}{Benett Axtell} {and} \bibinfo{person}{Cosmin Munteanu}.} \bibinfo{year}{2019}\natexlab{b}.
\newblock \showarticletitle{{PhotoFlow} in {Action}: {Picture}-{Mediated} {Reminiscence} {Supporting} {Family} {Socio}-{Connectivity}}. In \bibinfo{booktitle}{\emph{Extended {Abstracts} of the 2019 {CHI} {Conference} on {Human} {Factors} in {Computing} {Systems}}}. \bibinfo{publisher}{ACM}, \bibinfo{address}{Glasgow Scotland Uk}, \bibinfo{pages}{1--4}.
\newblock
\showISBNx{978-1-4503-5971-9}
\urldef\tempurl%
\url{https://doi.org/10.1145/3290607.3313272}
\showDOI{\tempurl}


\bibitem[Axtell et~al\mbox{.}(2022)]%
        {axtell_design_2022}
\bibfield{author}{\bibinfo{person}{Benett Axtell}, \bibinfo{person}{Raheleh Saryazdi}, {and} \bibinfo{person}{Cosmin Munteanu}.} \bibinfo{year}{2022}\natexlab{}.
\newblock \showarticletitle{Design is {Worth} a {Thousand} {Words}: {The} {Effect} of {Digital} {Interaction} {Design} on {Picture}-{Prompted} {Reminiscence}}. In \bibinfo{booktitle}{\emph{{CHI} {Conference} on {Human} {Factors} in {Computing} {Systems}}}. \bibinfo{publisher}{ACM}, \bibinfo{address}{New Orleans LA USA}, \bibinfo{pages}{1--12}.
\newblock
\showISBNx{978-1-4503-9157-3}
\urldef\tempurl%
\url{https://doi.org/10.1145/3491102.3517692}
\showDOI{\tempurl}


\bibitem[Baker et~al\mbox{.}(2021)]%
        {baker_schools_2021}
\bibfield{author}{\bibinfo{person}{Steven Baker}, \bibinfo{person}{Ryan~M. Kelly}, \bibinfo{person}{Jenny Waycott}, \bibinfo{person}{Romina Carrasco}, \bibinfo{person}{Roger Bell}, \bibinfo{person}{Zaher Joukhadar}, \bibinfo{person}{Thuong Hoang}, \bibinfo{person}{Elizabeth Ozanne}, {and} \bibinfo{person}{Frank Vetere}.} \bibinfo{year}{2021}\natexlab{}.
\newblock \showarticletitle{School's {Back}: {Scaffolding} {Reminiscence} in {Social} {Virtual} {Reality} with {Older} {Adults}}.
\newblock \bibinfo{journal}{\emph{Proceedings of the ACM on Human-Computer Interaction}} \bibinfo{volume}{4}, \bibinfo{number}{CSCW3} (\bibinfo{date}{Jan.} \bibinfo{year}{2021}), \bibinfo{pages}{1--25}.
\newblock
\showISSN{2573-0142}
\urldef\tempurl%
\url{https://doi.org/10.1145/3434176}
\showDOI{\tempurl}


\bibitem[Bluck and Levine(1998)]%
        {bluck_reminiscence_1998}
\bibfield{author}{\bibinfo{person}{Susan Bluck} {and} \bibinfo{person}{Linda~J. Levine}.} \bibinfo{year}{1998}\natexlab{}.
\newblock \showarticletitle{Reminiscence as autobiographical memory: a catalyst for reminiscence theory development}.
\newblock \bibinfo{journal}{\emph{Ageing and Society}} \bibinfo{volume}{18}, \bibinfo{number}{2} (\bibinfo{date}{March} \bibinfo{year}{1998}), \bibinfo{pages}{185--208}.
\newblock
\showISSN{0144-686X, 1469-1779}
\urldef\tempurl%
\url{https://doi.org/10.1017/S0144686X98006862}
\showDOI{\tempurl}


\bibitem[Braun and Clarke(2006)]%
        {braun_using_2006}
\bibfield{author}{\bibinfo{person}{Virginia Braun} {and} \bibinfo{person}{Victoria Clarke}.} \bibinfo{year}{2006}\natexlab{}.
\newblock \showarticletitle{Using thematic analysis in psychology}.
\newblock \bibinfo{journal}{\emph{Qualitative Research in Psychology}} \bibinfo{volume}{3}, \bibinfo{number}{2} (\bibinfo{date}{Jan.} \bibinfo{year}{2006}), \bibinfo{pages}{77--101}.
\newblock
\showISSN{1478-0887, 1478-0895}
\urldef\tempurl%
\url{https://doi.org/10.1191/1478088706qp063oa}
\showDOI{\tempurl}


\bibitem[Brewer and Piper(2016)]%
        {brewer_tell_2016}
\bibfield{author}{\bibinfo{person}{Robin Brewer} {and} \bibinfo{person}{Anne~Marie Piper}.} \bibinfo{year}{2016}\natexlab{}.
\newblock \showarticletitle{"{Tell} {It} {Like} {It} {Really} {Is}": {A} {Case} of {Online} {Content} {Creation} and {Sharing} {Among} {Older} {Adult} {Bloggers}}. In \bibinfo{booktitle}{\emph{Proceedings of the 2016 {CHI} {Conference} on {Human} {Factors} in {Computing} {Systems}}}. \bibinfo{publisher}{ACM}, \bibinfo{address}{San Jose California USA}, \bibinfo{pages}{5529--5542}.
\newblock
\showISBNx{978-1-4503-3362-7}
\urldef\tempurl%
\url{https://doi.org/10.1145/2858036.2858379}
\showDOI{\tempurl}


\bibitem[Brewer and Jones(2015)]%
        {brewer_pinteresce_2015}
\bibfield{author}{\bibinfo{person}{Robin~N. Brewer} {and} \bibinfo{person}{Jasmine Jones}.} \bibinfo{year}{2015}\natexlab{}.
\newblock \showarticletitle{Pinteresce: {Exploring} {Reminiscence} as an {Incentive} to {Digital} {Reciprocity} for {Older} {Adults}}. In \bibinfo{booktitle}{\emph{Proceedings of the 18th {ACM} {Conference} {Companion} on {Computer} {Supported} {Cooperative} {Work} \& {Social} {Computing}}}. \bibinfo{publisher}{ACM}, \bibinfo{address}{Vancouver BC Canada}, \bibinfo{pages}{243--246}.
\newblock
\showISBNx{978-1-4503-2946-0}
\urldef\tempurl%
\url{https://doi.org/10.1145/2685553.2699017}
\showDOI{\tempurl}


\bibitem[Broekhuijsen et~al\mbox{.}(2017a)]%
        {broekhuijsen_design_2017}
\bibfield{author}{\bibinfo{person}{Mendel Broekhuijsen}, \bibinfo{person}{Elise Van Den~Hoven}, {and} \bibinfo{person}{Panos Markopoulos}.} \bibinfo{year}{2017}\natexlab{a}.
\newblock \showarticletitle{Design {Directions} for {Media}-{Supported} {Collocated} {Remembering} {Practices}}. In \bibinfo{booktitle}{\emph{Proceedings of the {Eleventh} {International} {Conference} on {Tangible}, {Embedded}, and {Embodied} {Interaction}}}. \bibinfo{publisher}{ACM}, \bibinfo{address}{Yokohama Japan}, \bibinfo{pages}{21--30}.
\newblock
\showISBNx{978-1-4503-4676-4}
\urldef\tempurl%
\url{https://doi.org/10.1145/3024969.3024996}
\showDOI{\tempurl}


\bibitem[Broekhuijsen et~al\mbox{.}(2017b)]%
        {broekhuijsen_photowork_2017}
\bibfield{author}{\bibinfo{person}{Mendel Broekhuijsen}, \bibinfo{person}{Elise Van Den~Hoven}, {and} \bibinfo{person}{Panos Markopoulos}.} \bibinfo{year}{2017}\natexlab{b}.
\newblock \showarticletitle{From {PhotoWork} to {PhotoUse}: exploring personal digital photo activities}.
\newblock \bibinfo{journal}{\emph{Behaviour \& Information Technology}} \bibinfo{volume}{36}, \bibinfo{number}{7} (\bibinfo{date}{July} \bibinfo{year}{2017}), \bibinfo{pages}{754--767}.
\newblock
\showISSN{0144-929X, 1362-3001}
\urldef\tempurl%
\url{https://doi.org/10.1080/0144929X.2017.1288266}
\showDOI{\tempurl}


\bibitem[Cappeliez and O'Rourke(2006)]%
        {cappeliez_empirical_2006}
\bibfield{author}{\bibinfo{person}{P. Cappeliez} {and} \bibinfo{person}{N. O'Rourke}.} \bibinfo{year}{2006}\natexlab{}.
\newblock \showarticletitle{Empirical {Validation} of a {Model} of {Reminiscence} and {Health} in {Later} {Life}}.
\newblock \bibinfo{journal}{\emph{The Journals of Gerontology Series B: Psychological Sciences and Social Sciences}} \bibinfo{volume}{61}, \bibinfo{number}{4} (\bibinfo{date}{July} \bibinfo{year}{2006}), \bibinfo{pages}{P237--P244}.
\newblock
\showISSN{1079-5014, 1758-5368}
\urldef\tempurl%
\url{https://doi.org/10.1093/geronb/61.4.P237}
\showDOI{\tempurl}


\bibitem[Chen et~al\mbox{.}(2023)]%
        {chen_exploring_2023}
\bibfield{author}{\bibinfo{person}{Amy Yo~Sue Chen}, \bibinfo{person}{William Odom}, \bibinfo{person}{Carman Neustaedter}, \bibinfo{person}{Ce Zhong}, {and} \bibinfo{person}{Henry Lin}.} \bibinfo{year}{2023}\natexlab{}.
\newblock \showarticletitle{Exploring {Memory}-{Oriented} {Interactions} with {Digital} {Photos} {In} and {Across} {Time}: {A} {Field} {Study} of {Chronoscope}}. In \bibinfo{booktitle}{\emph{Proceedings of the 2023 {CHI} {Conference} on {Human} {Factors} in {Computing} {Systems}}}. \bibinfo{publisher}{ACM}, \bibinfo{address}{Hamburg Germany}, \bibinfo{pages}{1--20}.
\newblock
\showISBNx{978-1-4503-9421-5}
\urldef\tempurl%
\url{https://doi.org/10.1145/3544548.3581012}
\showDOI{\tempurl}


\bibitem[{Committee on the Health and Medical Dimensions of Social Isolation and Loneliness in Older Adults} et~al\mbox{.}(2020)]%
        {committee_on_the_health_and_medical_dimensions_of_social_isolation_and_loneliness_in_older_adults_social_2020}
\bibfield{author}{\bibinfo{person}{{Committee on the Health and Medical Dimensions of Social Isolation and Loneliness in Older Adults}}, \bibinfo{person}{{Board on Health Sciences Policy}}, \bibinfo{person}{{Board on Behavioral, Cognitive, and Sensory Sciences}}, \bibinfo{person}{{Health and Medicine Division}}, \bibinfo{person}{{Division of Behavioral and Social Sciences and Education}}, {and} \bibinfo{person}{{National Academies of Sciences, Engineering, and Medicine}}.} \bibinfo{year}{2020}\natexlab{}.
\newblock \bibinfo{booktitle}{\emph{Social {Isolation} and {Loneliness} in {Older} {Adults}: {Opportunities} for the {Health} {Care} {System}}}.
\newblock \bibinfo{publisher}{National Academies Press}, \bibinfo{address}{Washington, D.C.}
\newblock
\showISBNx{978-0-309-67100-2}
\urldef\tempurl%
\url{https://doi.org/10.17226/25663}
\showDOI{\tempurl}
\newblock
\shownote{Pages: 25663}.


\bibitem[Conway(2009)]%
        {conway_episodic_2009}
\bibfield{author}{\bibinfo{person}{Martin~A. Conway}.} \bibinfo{year}{2009}\natexlab{}.
\newblock \showarticletitle{Episodic memories}.
\newblock \bibinfo{journal}{\emph{Neuropsychologia}} \bibinfo{volume}{47}, \bibinfo{number}{11} (\bibinfo{date}{Sept.} \bibinfo{year}{2009}), \bibinfo{pages}{2305--2313}.
\newblock
\showISSN{00283932}
\urldef\tempurl%
\url{https://doi.org/10.1016/j.neuropsychologia.2009.02.003}
\showDOI{\tempurl}


\bibitem[Cousins(1989)]%
        {cousins_culture_1989}
\bibfield{author}{\bibinfo{person}{Steven~D. Cousins}.} \bibinfo{year}{1989}\natexlab{}.
\newblock \showarticletitle{Culture and self-perception in {Japan} and the {United} {States}.}
\newblock \bibinfo{journal}{\emph{Journal of Personality and Social Psychology}} \bibinfo{volume}{56}, \bibinfo{number}{1} (\bibinfo{date}{Jan.} \bibinfo{year}{1989}), \bibinfo{pages}{124--131}.
\newblock
\showISSN{1939-1315, 0022-3514}
\urldef\tempurl%
\url{https://doi.org/10.1037/0022-3514.56.1.124}
\showDOI{\tempurl}


\bibitem[Cuadra et~al\mbox{.}(2023)]%
        {cuadra_designing_2023}
\bibfield{author}{\bibinfo{person}{Andrea Cuadra}, \bibinfo{person}{Jessica Bethune}, \bibinfo{person}{Rony Krell}, \bibinfo{person}{Alexa Lempel}, \bibinfo{person}{Katrin Hänsel}, \bibinfo{person}{Armin Shahrokni}, \bibinfo{person}{Deborah Estrin}, {and} \bibinfo{person}{Nicola Dell}.} \bibinfo{year}{2023}\natexlab{}.
\newblock \showarticletitle{Designing {Voice}-{First} {Ambient} {Interfaces} to {Support} {Aging} in {Place}}. In \bibinfo{booktitle}{\emph{Proceedings of the 2023 {ACM} {Designing} {Interactive} {Systems} {Conference}}}. \bibinfo{publisher}{ACM}, \bibinfo{address}{Pittsburgh PA USA}, \bibinfo{pages}{2189--2205}.
\newblock
\showISBNx{978-1-4503-9893-0}
\urldef\tempurl%
\url{https://doi.org/10.1145/3563657.3596104}
\showDOI{\tempurl}


\bibitem[Davidson and Jensen(2013)]%
        {davidson_participatory_2013}
\bibfield{author}{\bibinfo{person}{Jennifer~L. Davidson} {and} \bibinfo{person}{Carlos Jensen}.} \bibinfo{year}{2013}\natexlab{}.
\newblock \showarticletitle{Participatory design with older adults: an analysis of creativity in the design of mobile healthcare applications}. In \bibinfo{booktitle}{\emph{Proceedings of the 9th {ACM} {Conference} on {Creativity} \& {Cognition}}}. \bibinfo{publisher}{ACM}, \bibinfo{address}{Sydney Australia}, \bibinfo{pages}{114--123}.
\newblock
\showISBNx{978-1-4503-2150-1}
\urldef\tempurl%
\url{https://doi.org/10.1145/2466627.2466652}
\showDOI{\tempurl}


\bibitem[Ding et~al\mbox{.}(2022)]%
        {ding_talktive_2022}
\bibfield{author}{\bibinfo{person}{Zijian Ding}, \bibinfo{person}{Jiawen Kang}, \bibinfo{person}{Tinky Oi~Ting Ho}, \bibinfo{person}{Ka~Ho Wong}, \bibinfo{person}{Helene~H Fung}, \bibinfo{person}{Helen Meng}, {and} \bibinfo{person}{Xiaojuan Ma}.} \bibinfo{year}{2022}\natexlab{}.
\newblock \showarticletitle{{TalkTive}: {A} {Conversational} {Agent} {Using} {Backchannels} to {Engage} {Older} {Adults} in {Neurocognitive} {Disorders} {Screening}}. In \bibinfo{booktitle}{\emph{{CHI} {Conference} on {Human} {Factors} in {Computing} {Systems}}}. \bibinfo{publisher}{ACM}, \bibinfo{address}{New Orleans LA USA}, \bibinfo{pages}{1--19}.
\newblock
\showISBNx{978-1-4503-9157-3}
\urldef\tempurl%
\url{https://doi.org/10.1145/3491102.3502005}
\showDOI{\tempurl}


\bibitem[El~Haj et~al\mbox{.}(2020)]%
        {el_haj_picture_2020}
\bibfield{author}{\bibinfo{person}{Mohamad El~Haj}, \bibinfo{person}{Dimitrios Kapogiannis}, {and} \bibinfo{person}{Pascal Antoine}.} \bibinfo{year}{2020}\natexlab{}.
\newblock \showarticletitle{The picture of the past: {Pictures} to cue autobiographical memory in {Alzheimer}’s disease}.
\newblock \bibinfo{journal}{\emph{Journal of Clinical and Experimental Neuropsychology}} \bibinfo{volume}{42}, \bibinfo{number}{9} (\bibinfo{date}{Oct.} \bibinfo{year}{2020}), \bibinfo{pages}{914--923}.
\newblock
\showISSN{1380-3395, 1744-411X}
\urldef\tempurl%
\url{https://doi.org/10.1080/13803395.2020.1825636}
\showDOI{\tempurl}


\bibitem[Frohlich et~al\mbox{.}(2002)]%
        {frohlich_requirements_2002}
\bibfield{author}{\bibinfo{person}{David Frohlich}, \bibinfo{person}{Allan Kuchinsky}, \bibinfo{person}{Celine Pering}, \bibinfo{person}{Abbe Don}, {and} \bibinfo{person}{Steven Ariss}.} \bibinfo{year}{2002}\natexlab{}.
\newblock \showarticletitle{Requirements for photoware}. In \bibinfo{booktitle}{\emph{Proceedings of the 2002 {ACM} conference on {Computer} supported cooperative work}}. \bibinfo{publisher}{ACM}, \bibinfo{address}{New Orleans Louisiana USA}, \bibinfo{pages}{166--175}.
\newblock
\showISBNx{978-1-58113-560-2}
\urldef\tempurl%
\url{https://doi.org/10.1145/587078.587102}
\showDOI{\tempurl}


\bibitem[Graham et~al\mbox{.}(2002)]%
        {graham_time_2002}
\bibfield{author}{\bibinfo{person}{Adrian Graham}, \bibinfo{person}{Hector Garcia-Molina}, \bibinfo{person}{Andreas Paepcke}, {and} \bibinfo{person}{Terry Winograd}.} \bibinfo{year}{2002}\natexlab{}.
\newblock \showarticletitle{Time as essence for photo browsing through personal digital libraries}. In \bibinfo{booktitle}{\emph{Proceedings of the 2nd {ACM}/{IEEE}-{CS} joint conference on {Digital} libraries}}. \bibinfo{publisher}{ACM}, \bibinfo{address}{Portland Oregon USA}, \bibinfo{pages}{326--335}.
\newblock
\showISBNx{978-1-58113-513-8}
\urldef\tempurl%
\url{https://doi.org/10.1145/544220.544301}
\showDOI{\tempurl}


\bibitem[Greenberg et~al\mbox{.}(2005)]%
        {greenberg_visual_2005}
\bibfield{author}{\bibinfo{person}{D Greenberg}, \bibinfo{person}{M Eacott}, \bibinfo{person}{D Brechin}, {and} \bibinfo{person}{D Rubin}.} \bibinfo{year}{2005}\natexlab{}.
\newblock \showarticletitle{Visual memory loss and autobiographical amnesia: a case study}.
\newblock \bibinfo{journal}{\emph{Neuropsychologia}} \bibinfo{volume}{43}, \bibinfo{number}{10} (\bibinfo{year}{2005}), \bibinfo{pages}{1493--1502}.
\newblock
\showISSN{00283932}
\urldef\tempurl%
\url{https://doi.org/10.1016/j.neuropsychologia.2004.12.009}
\showDOI{\tempurl}


\bibitem[Istvandity(2017)]%
        {istvandity_combining_2017}
\bibfield{author}{\bibinfo{person}{Lauren Istvandity}.} \bibinfo{year}{2017}\natexlab{}.
\newblock \showarticletitle{Combining music and reminiscence therapy interventions for wellbeing in elderly populations: {A} systematic review}.
\newblock \bibinfo{journal}{\emph{Complementary Therapies in Clinical Practice}}  \bibinfo{volume}{28} (\bibinfo{date}{Aug.} \bibinfo{year}{2017}), \bibinfo{pages}{18--25}.
\newblock
\showISSN{17443881}
\urldef\tempurl%
\url{https://doi.org/10.1016/j.ctcp.2017.03.003}
\showDOI{\tempurl}


\bibitem[Ito et~al\mbox{.}(2006)]%
        {ito_personal_2006}
\bibfield{editor}{\bibinfo{person}{Mizuko Ito}, \bibinfo{person}{Misa Matsuda}, {and} \bibinfo{person}{Daisuke Okabe}} (Eds.). \bibinfo{year}{2006}\natexlab{}.
\newblock \bibinfo{booktitle}{\emph{Personal, {Portable}, {Pedestrian}: {Mobile} {Phones} in {Japanese} {Life}}}.
\newblock \bibinfo{publisher}{The MIT Press}.
\newblock
\showISBNx{978-0-262-25641-4}
\urldef\tempurl%
\url{https://doi.org/10.7551/mitpress/5309.001.0001}
\showDOI{\tempurl}


\bibitem[Jin et~al\mbox{.}(2024)]%
        {jin_exploring_2024}
\bibfield{author}{\bibinfo{person}{Yucheng Jin}, \bibinfo{person}{Wanling Cai}, \bibinfo{person}{Li Chen}, \bibinfo{person}{Yizhe Zhang}, \bibinfo{person}{Gavin Doherty}, {and} \bibinfo{person}{Tonglin Jiang}.} \bibinfo{year}{2024}\natexlab{}.
\newblock \showarticletitle{Exploring the {Design} of {Generative} {AI} in {Supporting} {Music}-based {Reminiscence} for {Older} {Adults}}. In \bibinfo{booktitle}{\emph{Proceedings of the {CHI} {Conference} on {Human} {Factors} in {Computing} {Systems}}}. \bibinfo{publisher}{ACM}, \bibinfo{address}{Honolulu HI USA}, \bibinfo{pages}{1--17}.
\newblock
\showISBNx{9798400703300}
\urldef\tempurl%
\url{https://doi.org/10.1145/3613904.3642800}
\showDOI{\tempurl}


\bibitem[Jones and Ackerman(2018)]%
        {jones_co-constructing_2018-1}
\bibfield{author}{\bibinfo{person}{Jasmine Jones} {and} \bibinfo{person}{Mark~S. Ackerman}.} \bibinfo{year}{2018}\natexlab{}.
\newblock \showarticletitle{Co-constructing {Family} {Memory}: {Understanding} the {Intergenerational} {Practices} of {Passing} on {Family} {Stories}}. In \bibinfo{booktitle}{\emph{Proceedings of the 2018 {CHI} {Conference} on {Human} {Factors} in {Computing} {Systems}}}. \bibinfo{publisher}{ACM}, \bibinfo{address}{Montreal QC Canada}, \bibinfo{pages}{1--13}.
\newblock
\showISBNx{978-1-4503-5620-6}
\urldef\tempurl%
\url{https://doi.org/10.1145/3173574.3173998}
\showDOI{\tempurl}


\bibitem[Kang et~al\mbox{.}(2021)]%
        {kang_momentmeld_2021}
\bibfield{author}{\bibinfo{person}{Bumsoo Kang}, \bibinfo{person}{Seungwoo Kang}, {and} \bibinfo{person}{Inseok Hwang}.} \bibinfo{year}{2021}\natexlab{}.
\newblock \showarticletitle{{MomentMeld}: {AI}-augmented {Mobile} {Photographic} {Memento} towards {Mutually} {Stimulatory} {Inter}-generational {Interaction}}. In \bibinfo{booktitle}{\emph{Proceedings of the 2021 {CHI} {Conference} on {Human} {Factors} in {Computing} {Systems}}}. \bibinfo{publisher}{ACM}, \bibinfo{address}{Yokohama Japan}, \bibinfo{pages}{1--16}.
\newblock
\showISBNx{978-1-4503-8096-6}
\urldef\tempurl%
\url{https://doi.org/10.1145/3411764.3445688}
\showDOI{\tempurl}


\bibitem[Keightley and Pickering(2014)]%
        {keightley_technologies_2014}
\bibfield{author}{\bibinfo{person}{Emily Keightley} {and} \bibinfo{person}{Michael Pickering}.} \bibinfo{year}{2014}\natexlab{}.
\newblock \showarticletitle{Technologies of memory: {Practices} of remembering in analogue and digital photography}.
\newblock \bibinfo{journal}{\emph{New Media \& Society}} \bibinfo{volume}{16}, \bibinfo{number}{4} (\bibinfo{date}{June} \bibinfo{year}{2014}), \bibinfo{pages}{576--593}.
\newblock
\showISSN{1461-4448, 1461-7315}
\urldef\tempurl%
\url{https://doi.org/10.1177/1461444814532062}
\showDOI{\tempurl}


\bibitem[Kim et~al\mbox{.}(2011)]%
        {kim_photo_2011}
\bibfield{author}{\bibinfo{person}{Heung-Nam Kim}, \bibinfo{person}{Abdulmotaleb El~Saddik}, \bibinfo{person}{Kee-Sung Lee}, \bibinfo{person}{Yeon-Ho Lee}, {and} \bibinfo{person}{Geun-Sik Jo}.} \bibinfo{year}{2011}\natexlab{}.
\newblock \showarticletitle{Photo search in a personal photo diary by drawing face position with people tagging}. In \bibinfo{booktitle}{\emph{Proceedings of the 16th international conference on {Intelligent} user interfaces}}. \bibinfo{publisher}{ACM}, \bibinfo{address}{Palo Alto CA USA}, \bibinfo{pages}{443--444}.
\newblock
\showISBNx{978-1-4503-0419-1}
\urldef\tempurl%
\url{https://doi.org/10.1145/1943403.1943490}
\showDOI{\tempurl}


\bibitem[Kim et~al\mbox{.}(2022)]%
        {kim_slide2remember_2022}
\bibfield{author}{\bibinfo{person}{Subin Kim}, \bibinfo{person}{Sangsu Jang}, \bibinfo{person}{Jin-young Moon}, \bibinfo{person}{Minjoo Han}, {and} \bibinfo{person}{Young-Woo Park}.} \bibinfo{year}{2022}\natexlab{}.
\newblock \showarticletitle{{Slide2Remember}: an {Interactive} {Wall} {Frame} {Enriching} {Reminiscence} {Experiences} by {Providing} {Re}-encounters of {Taken} {Photos} and {Heard} {Music} in a {Similar} {Period}}. In \bibinfo{booktitle}{\emph{Designing {Interactive} {Systems} {Conference}}}. \bibinfo{publisher}{ACM}, \bibinfo{address}{Virtual Event Australia}, \bibinfo{pages}{288--300}.
\newblock
\showISBNx{978-1-4503-9358-4}
\urldef\tempurl%
\url{https://doi.org/10.1145/3532106.3533456}
\showDOI{\tempurl}


\bibitem[Kirk et~al\mbox{.}(2006)]%
        {kirk_understanding_2006}
\bibfield{author}{\bibinfo{person}{David Kirk}, \bibinfo{person}{Abigail Sellen}, \bibinfo{person}{Carsten Rother}, {and} \bibinfo{person}{Ken Wood}.} \bibinfo{year}{2006}\natexlab{}.
\newblock \showarticletitle{Understanding photowork}. In \bibinfo{booktitle}{\emph{Proceedings of the {SIGCHI} {Conference} on {Human} {Factors} in {Computing} {Systems}}}. \bibinfo{publisher}{ACM}, \bibinfo{address}{Montréal Québec Canada}, \bibinfo{pages}{761--770}.
\newblock
\showISBNx{978-1-59593-372-0}
\urldef\tempurl%
\url{https://doi.org/10.1145/1124772.1124885}
\showDOI{\tempurl}


\bibitem[Lazar et~al\mbox{.}(2014)]%
        {lazar_systematic_2014}
\bibfield{author}{\bibinfo{person}{Amanda Lazar}, \bibinfo{person}{Hilaire Thompson}, {and} \bibinfo{person}{George Demiris}.} \bibinfo{year}{2014}\natexlab{}.
\newblock \showarticletitle{A {Systematic} {Review} of the {Use} of {Technology} for {Reminiscence} {Therapy}}.
\newblock \bibinfo{journal}{\emph{Health Education \& Behavior}} \bibinfo{volume}{41}, \bibinfo{number}{1\_suppl} (\bibinfo{date}{Oct.} \bibinfo{year}{2014}), \bibinfo{pages}{51S--61S}.
\newblock
\showISSN{1090-1981, 1552-6127}
\urldef\tempurl%
\url{https://doi.org/10.1177/1090198114537067}
\showDOI{\tempurl}


\bibitem[Lee et~al\mbox{.}(2014)]%
        {lee_picgo_2014}
\bibfield{author}{\bibinfo{person}{Hung-Chi Lee}, \bibinfo{person}{Ya~Fang Cheng}, \bibinfo{person}{Szu~Yang Cho}, \bibinfo{person}{Hsien-Hui Tang}, \bibinfo{person}{Jane Hsu}, {and} \bibinfo{person}{Chien-Hsiung Chen}.} \bibinfo{year}{2014}\natexlab{}.
\newblock \showarticletitle{Picgo: designing reminiscence and storytelling for the elderly with photo annotation}. In \bibinfo{booktitle}{\emph{Proceedings of the 2014 companion publication on {Designing} interactive systems}}. \bibinfo{publisher}{ACM}, \bibinfo{address}{Vancouver BC Canada}, \bibinfo{pages}{9--12}.
\newblock
\showISBNx{978-1-4503-2903-3}
\urldef\tempurl%
\url{https://doi.org/10.1145/2598784.2602769}
\showDOI{\tempurl}


\bibitem[Lee and Hsu(2016)]%
        {lee_picmemory_2016}
\bibfield{author}{\bibinfo{person}{Hung-Chi Lee} {and} \bibinfo{person}{Jane Yung-jen Hsu}.} \bibinfo{year}{2016}\natexlab{}.
\newblock \showarticletitle{{PicMemory}: {Enriching} {Intergenerational} {Family} {Interaction} and {Memory} {Collection}}. In \bibinfo{booktitle}{\emph{Proceedings of the 2016 {CHI} {Conference} {Extended} {Abstracts} on {Human} {Factors} in {Computing} {Systems}}}. \bibinfo{publisher}{ACM}, \bibinfo{address}{San Jose California USA}, \bibinfo{pages}{3715--3718}.
\newblock
\showISBNx{978-1-4503-4082-3}
\urldef\tempurl%
\url{https://doi.org/10.1145/2851581.2890233}
\showDOI{\tempurl}


\bibitem[Lee and Dey(2007)]%
        {lee_providing_2007}
\bibfield{author}{\bibinfo{person}{Matthew~L. Lee} {and} \bibinfo{person}{Anind~K. Dey}.} \bibinfo{year}{2007}\natexlab{}.
\newblock \showarticletitle{Providing good memory cues for people with episodic memory impairment}. In \bibinfo{booktitle}{\emph{Proceedings of the 9th international {ACM} {SIGACCESS} conference on {Computers} and accessibility}}. \bibinfo{publisher}{ACM}, \bibinfo{address}{Tempe Arizona USA}, \bibinfo{pages}{131--138}.
\newblock
\showISBNx{978-1-59593-573-1}
\urldef\tempurl%
\url{https://doi.org/10.1145/1296843.1296867}
\showDOI{\tempurl}


\bibitem[Li et~al\mbox{.}(2019)]%
        {li_measuring_2019}
\bibfield{author}{\bibinfo{person}{Jie Li}, \bibinfo{person}{Yiping Kong}, \bibinfo{person}{Thomas Röggla}, \bibinfo{person}{Francesca De~Simone}, \bibinfo{person}{Swamy Ananthanarayan}, \bibinfo{person}{Huib De~Ridder}, \bibinfo{person}{Abdallah El~Ali}, {and} \bibinfo{person}{Pablo Cesar}.} \bibinfo{year}{2019}\natexlab{}.
\newblock \showarticletitle{Measuring and {Understanding} {Photo} {Sharing} {Experiences} in {Social} {Virtual} {Reality}}. In \bibinfo{booktitle}{\emph{Proceedings of the 2019 {CHI} {Conference} on {Human} {Factors} in {Computing} {Systems}}}. \bibinfo{publisher}{ACM}, \bibinfo{address}{Glasgow Scotland Uk}, \bibinfo{pages}{1--14}.
\newblock
\showISBNx{978-1-4503-5970-2}
\urldef\tempurl%
\url{https://doi.org/10.1145/3290605.3300897}
\showDOI{\tempurl}


\bibitem[Li et~al\mbox{.}(2023)]%
        {li_exploring_2023}
\bibfield{author}{\bibinfo{person}{Zisu Li}, \bibinfo{person}{Li Feng}, \bibinfo{person}{Chen Liang}, \bibinfo{person}{Yuru Huang}, {and} \bibinfo{person}{Mingming Fan}.} \bibinfo{year}{2023}\natexlab{}.
\newblock \showarticletitle{Exploring the {Opportunities} of {AR} for {Enriching} {Storytelling} with {Family} {Photos} between {Grandparents} and {Grandchildren}}.
\newblock \bibinfo{journal}{\emph{Proceedings of the ACM on Interactive, Mobile, Wearable and Ubiquitous Technologies}} \bibinfo{volume}{7}, \bibinfo{number}{3} (\bibinfo{date}{Sept.} \bibinfo{year}{2023}), \bibinfo{pages}{1--26}.
\newblock
\showISSN{2474-9567}
\urldef\tempurl%
\url{https://doi.org/10.1145/3610903}
\showDOI{\tempurl}


\bibitem[Lindley and Wallace(2015)]%
        {lindley_placing_2015}
\bibfield{author}{\bibinfo{person}{Siân Lindley} {and} \bibinfo{person}{Jayne Wallace}.} \bibinfo{year}{2015}\natexlab{}.
\newblock \showarticletitle{Placing in {Age}: {Transitioning} to a {New} {Home} in {Later} {Life}}.
\newblock \bibinfo{journal}{\emph{ACM Transactions on Computer-Human Interaction}} \bibinfo{volume}{22}, \bibinfo{number}{4} (\bibinfo{date}{July} \bibinfo{year}{2015}), \bibinfo{pages}{1--39}.
\newblock
\showISSN{1073-0516, 1557-7325}
\urldef\tempurl%
\url{https://doi.org/10.1145/2755562}
\showDOI{\tempurl}


\bibitem[Lindley(2012)]%
        {lindley_before_2012}
\bibfield{author}{\bibinfo{person}{Siân~E. Lindley}.} \bibinfo{year}{2012}\natexlab{}.
\newblock \showarticletitle{Before {I} {Forget}: {From} {Personal} {Memory} to {Family} {History}}.
\newblock \bibinfo{journal}{\emph{Human–Computer Interaction}}  \bibinfo{volume}{27} (\bibinfo{year}{2012}), \bibinfo{pages}{13--36}.
\newblock
\urldef\tempurl%
\url{https://doi.org/10.1080/07370024.2012.656065}
\showDOI{\tempurl}


\bibitem[McAdams et~al\mbox{.}(2001)]%
        {mcadams_when_2001}
\bibfield{author}{\bibinfo{person}{Dan~P. McAdams}, \bibinfo{person}{Jeffrey Reynolds}, \bibinfo{person}{Martha Lewis}, \bibinfo{person}{Allison~H. Patten}, {and} \bibinfo{person}{Phillip~J. Bowman}.} \bibinfo{year}{2001}\natexlab{}.
\newblock \showarticletitle{When {Bad} {Things} {Turn} {Good} and {Good} {Things} {Turn} {Bad}: {Sequences} of {Redemption} and {Contamination} in {Life} {Narrative} and their {Relation} to {Psychosocial} {Adaptation} in {Midlife} {Adults} and in {Students}}.
\newblock \bibinfo{journal}{\emph{Personality and Social Psychology Bulletin}} \bibinfo{volume}{27}, \bibinfo{number}{4} (\bibinfo{date}{April} \bibinfo{year}{2001}), \bibinfo{pages}{474--485}.
\newblock
\showISSN{0146-1672, 1552-7433}
\urldef\tempurl%
\url{https://doi.org/10.1177/0146167201274008}
\showDOI{\tempurl}


\bibitem[McGookin(2019)]%
        {mcgookin_reveal_2019}
\bibfield{author}{\bibinfo{person}{David McGookin}.} \bibinfo{year}{2019}\natexlab{}.
\newblock \showarticletitle{Reveal: {Investigating} {Proactive} {Location}-{Based} {Reminiscing} with {Personal} {Digital} {Photo} {Repositories}}. In \bibinfo{booktitle}{\emph{Proceedings of the 2019 {CHI} {Conference} on {Human} {Factors} in {Computing} {Systems}}}. \bibinfo{publisher}{ACM}, \bibinfo{address}{Glasgow Scotland Uk}, \bibinfo{pages}{1--14}.
\newblock
\showISBNx{978-1-4503-5970-2}
\urldef\tempurl%
\url{https://doi.org/10.1145/3290605.3300665}
\showDOI{\tempurl}


\bibitem[Mongrain and Anselmo-Matthews(2012)]%
        {mongrain_positive_2012}
\bibfield{author}{\bibinfo{person}{Myriam Mongrain} {and} \bibinfo{person}{Tracy Anselmo-Matthews}.} \bibinfo{year}{2012}\natexlab{}.
\newblock \showarticletitle{Do {Positive} {Psychology} {Exercises} {Work}? {A} {Replication} of {Seligman} et al. ()}.
\newblock \bibinfo{journal}{\emph{Journal of Clinical Psychology}} \bibinfo{volume}{68}, \bibinfo{number}{4} (\bibinfo{date}{April} \bibinfo{year}{2012}), \bibinfo{pages}{jclp.21839}.
\newblock
\showISSN{0021-9762, 1097-4679}
\urldef\tempurl%
\url{https://doi.org/10.1002/jclp.21839}
\showDOI{\tempurl}


\bibitem[Petrelli et~al\mbox{.}(2009)]%
        {petrelli_making_2009}
\bibfield{author}{\bibinfo{person}{Daniela Petrelli}, \bibinfo{person}{Elise Van Den~Hoven}, {and} \bibinfo{person}{Steve Whittaker}.} \bibinfo{year}{2009}\natexlab{}.
\newblock \showarticletitle{Making history: intentional capture of future memories}. In \bibinfo{booktitle}{\emph{Proceedings of the {SIGCHI} {Conference} on {Human} {Factors} in {Computing} {Systems}}}. \bibinfo{publisher}{ACM}, \bibinfo{address}{Boston MA USA}, \bibinfo{pages}{1723--1732}.
\newblock
\showISBNx{978-1-60558-246-7}
\urldef\tempurl%
\url{https://doi.org/10.1145/1518701.1518966}
\showDOI{\tempurl}


\bibitem[Piper et~al\mbox{.}(2013)]%
        {piper_audio-enhanced_2013}
\bibfield{author}{\bibinfo{person}{Anne~Marie Piper}, \bibinfo{person}{Nadir Weibel}, {and} \bibinfo{person}{James Hollan}.} \bibinfo{year}{2013}\natexlab{}.
\newblock \showarticletitle{Audio-enhanced paper photos: encouraging social interaction at age 105}. In \bibinfo{booktitle}{\emph{Proceedings of the 2013 conference on {Computer} supported cooperative work}}. \bibinfo{publisher}{ACM}, \bibinfo{address}{San Antonio Texas USA}, \bibinfo{pages}{215--224}.
\newblock
\showISBNx{978-1-4503-1331-5}
\urldef\tempurl%
\url{https://doi.org/10.1145/2441776.2441802}
\showDOI{\tempurl}


\bibitem[Qian and Feijs(2004)]%
        {qian_exploring_2004}
\bibfield{author}{\bibinfo{person}{Yuechen Qian} {and} \bibinfo{person}{Loe M.~G. Feijs}.} \bibinfo{year}{2004}\natexlab{}.
\newblock \showarticletitle{Exploring the potentials of combining photo annotating tasks with instant messaging fun}. In \bibinfo{booktitle}{\emph{Proceedings of the 3rd international conference on {Mobile} and ubiquitous multimedia}}. \bibinfo{publisher}{ACM}, \bibinfo{address}{College Park Maryland USA}, \bibinfo{pages}{11--17}.
\newblock
\showISBNx{978-1-58113-981-5}
\urldef\tempurl%
\url{https://doi.org/10.1145/1052380.1052383}
\showDOI{\tempurl}


\bibitem[Rodden and Wood(2003)]%
        {rodden_how_2003}
\bibfield{author}{\bibinfo{person}{Kerry Rodden} {and} \bibinfo{person}{Kenneth~R. Wood}.} \bibinfo{year}{2003}\natexlab{}.
\newblock \showarticletitle{How do people manage their digital photographs?}. In \bibinfo{booktitle}{\emph{Proceedings of the {SIGCHI} {Conference} on {Human} {Factors} in {Computing} {Systems}}}. \bibinfo{publisher}{ACM}, \bibinfo{address}{Ft. Lauderdale Florida USA}, \bibinfo{pages}{409--416}.
\newblock
\showISBNx{978-1-58113-630-2}
\urldef\tempurl%
\url{https://doi.org/10.1145/642611.642682}
\showDOI{\tempurl}


\bibitem[Rogers et~al\mbox{.}(2014)]%
        {rogers_never_2014}
\bibfield{author}{\bibinfo{person}{Yvonne Rogers}, \bibinfo{person}{Jeni Paay}, \bibinfo{person}{Margot Brereton}, \bibinfo{person}{Kate~L. Vaisutis}, \bibinfo{person}{Gary Marsden}, {and} \bibinfo{person}{Frank Vetere}.} \bibinfo{year}{2014}\natexlab{}.
\newblock \showarticletitle{Never too old: engaging retired people inventing the future with {MaKey} {MaKey}}. In \bibinfo{booktitle}{\emph{Proceedings of the {SIGCHI} {Conference} on {Human} {Factors} in {Computing} {Systems}}}. \bibinfo{publisher}{ACM}, \bibinfo{address}{Toronto Ontario Canada}, \bibinfo{pages}{3913--3922}.
\newblock
\showISBNx{978-1-4503-2473-1}
\urldef\tempurl%
\url{https://doi.org/10.1145/2556288.2557184}
\showDOI{\tempurl}


\bibitem[Rubin(2005)]%
        {rubin_basic-systems_2005}
\bibfield{author}{\bibinfo{person}{David~C. Rubin}.} \bibinfo{year}{2005}\natexlab{}.
\newblock \showarticletitle{A {Basic}-{Systems} {Approach} to {Autobiographical} {Memory}}.
\newblock \bibinfo{journal}{\emph{Current Directions in Psychological Science}} \bibinfo{volume}{14}, \bibinfo{number}{2} (\bibinfo{date}{April} \bibinfo{year}{2005}), \bibinfo{pages}{79--83}.
\newblock
\showISSN{0963-7214, 1467-8721}
\urldef\tempurl%
\url{https://doi.org/10.1111/j.0963-7214.2005.00339.x}
\showDOI{\tempurl}


\bibitem[Sas(2018)]%
        {sas_exploring_2018}
\bibfield{author}{\bibinfo{person}{Corina Sas}.} \bibinfo{year}{2018}\natexlab{}.
\newblock \showarticletitle{Exploring {Self}-{Defining} {Memories} in {Old} {Age} and their {Digital} {Cues}}. In \bibinfo{booktitle}{\emph{Proceedings of the 2018 {Designing} {Interactive} {Systems} {Conference}}}. \bibinfo{publisher}{ACM}, \bibinfo{address}{Hong Kong China}, \bibinfo{pages}{149--161}.
\newblock
\showISBNx{978-1-4503-5198-0}
\urldef\tempurl%
\url{https://doi.org/10.1145/3196709.3196767}
\showDOI{\tempurl}


\bibitem[Simpson et~al\mbox{.}(2020)]%
        {simpson_daisy_2020}
\bibfield{author}{\bibinfo{person}{James Simpson}, \bibinfo{person}{Franziska Gaiser}, \bibinfo{person}{Miroslav Macík}, {and} \bibinfo{person}{Timna Breßgott}.} \bibinfo{year}{2020}\natexlab{}.
\newblock \showarticletitle{Daisy: {A} {Friendly} {Conversational} {Agent} for {Older} {Adults}}. In \bibinfo{booktitle}{\emph{Proceedings of the 2nd {Conference} on {Conversational} {User} {Interfaces}}}. \bibinfo{publisher}{ACM}, \bibinfo{address}{Bilbao Spain}, \bibinfo{pages}{1--3}.
\newblock
\showISBNx{978-1-4503-7544-3}
\urldef\tempurl%
\url{https://doi.org/10.1145/3405755.3406166}
\showDOI{\tempurl}


\bibitem[Sit et~al\mbox{.}(2005)]%
        {sit_digital_2005}
\bibfield{author}{\bibinfo{person}{R.Y. Sit}, \bibinfo{person}{J.D. Hollan}, {and} \bibinfo{person}{W.G. Griswold}.} \bibinfo{year}{2005}\natexlab{}.
\newblock \showarticletitle{Digital {Photos} as {Conversational} {Anchors}}. In \bibinfo{booktitle}{\emph{Proceedings of the 38th {Annual} {Hawaii} {International} {Conference} on {System} {Sciences}}}. \bibinfo{publisher}{IEEE}, \bibinfo{address}{Big Island, HI, USA}, \bibinfo{pages}{109b--109b}.
\newblock
\showISBNx{978-0-7695-2268-5}
\urldef\tempurl%
\url{https://doi.org/10.1109/HICSS.2005.203}
\showDOI{\tempurl}


\bibitem[Staudinger(2001)]%
        {staudinger_life_2001}
\bibfield{author}{\bibinfo{person}{Ursula~M. Staudinger}.} \bibinfo{year}{2001}\natexlab{}.
\newblock \showarticletitle{Life {Reflection}: {A} {Social}–{Cognitive} {Analysis} of {Life} {Review}}.
\newblock \bibinfo{journal}{\emph{Review of General Psychology}} \bibinfo{volume}{5}, \bibinfo{number}{2} (\bibinfo{date}{June} \bibinfo{year}{2001}), \bibinfo{pages}{148--160}.
\newblock
\showISSN{1089-2680, 1939-1552}
\urldef\tempurl%
\url{https://doi.org/10.1037/1089-2680.5.2.148}
\showDOI{\tempurl}


\bibitem[Tam et~al\mbox{.}(2021)]%
        {tam_effectiveness_2021}
\bibfield{author}{\bibinfo{person}{Wilson Tam}, \bibinfo{person}{Sum~Nok Poon}, \bibinfo{person}{Rathi Mahendran}, \bibinfo{person}{Ee~Heok Kua}, {and} \bibinfo{person}{Xi~Vivien Wu}.} \bibinfo{year}{2021}\natexlab{}.
\newblock \showarticletitle{The effectiveness of reminiscence-based intervention on improving psychological well-being in cognitively intact older adults: {A} systematic review and meta-analysis}.
\newblock \bibinfo{journal}{\emph{International Journal of Nursing Studies}}  \bibinfo{volume}{114} (\bibinfo{date}{Feb.} \bibinfo{year}{2021}), \bibinfo{pages}{103847}.
\newblock
\showISSN{00207489}
\urldef\tempurl%
\url{https://doi.org/10.1016/j.ijnurstu.2020.103847}
\showDOI{\tempurl}


\bibitem[Tang et~al\mbox{.}(2007)]%
        {tang_memory_2007}
\bibfield{author}{\bibinfo{person}{Karen~P. Tang}, \bibinfo{person}{Jason~I. Hong}, \bibinfo{person}{Ian~E. Smith}, \bibinfo{person}{Annie Ha}, {and} \bibinfo{person}{Lalatendu Satpathy}.} \bibinfo{year}{2007}\natexlab{}.
\newblock \showarticletitle{Memory karaoke: using a location-aware mobile reminiscence tool to support aging in place}. In \bibinfo{booktitle}{\emph{Proceedings of the 9th international conference on {Human} computer interaction with mobile devices and services}}. \bibinfo{publisher}{ACM}, \bibinfo{address}{Singapore}, \bibinfo{pages}{305--312}.
\newblock
\showISBNx{978-1-59593-862-6}
\urldef\tempurl%
\url{https://doi.org/10.1145/1377999.1378023}
\showDOI{\tempurl}


\bibitem[Tang et~al\mbox{.}(2023)]%
        {tang_towards_2023}
\bibfield{author}{\bibinfo{person}{Xinru Tang}, \bibinfo{person}{Xianghua~(Sharon) Ding}, {and} \bibinfo{person}{Zhixuan Zhou}.} \bibinfo{year}{2023}\natexlab{}.
\newblock \showarticletitle{Towards {Equitable} {Online} {Participation}: {A} {Case} of {Older} {Adult} {Content} {Creators}' {Role} {Transition} on {Short}-form {Video} {Sharing} {Platforms}}.
\newblock \bibinfo{journal}{\emph{Proceedings of the ACM on Human-Computer Interaction}} \bibinfo{volume}{7}, \bibinfo{number}{CSCW2} (\bibinfo{date}{Sept.} \bibinfo{year}{2023}), \bibinfo{pages}{1--22}.
\newblock
\showISSN{2573-0142}
\urldef\tempurl%
\url{https://doi.org/10.1145/3610216}
\showDOI{\tempurl}


\bibitem[Thiry et~al\mbox{.}(2013)]%
        {thiry_authoring_2013}
\bibfield{author}{\bibinfo{person}{Elizabeth Thiry}, \bibinfo{person}{Siân Lindley}, \bibinfo{person}{Richard Banks}, {and} \bibinfo{person}{Tim Regan}.} \bibinfo{year}{2013}\natexlab{}.
\newblock \showarticletitle{Authoring personal histories: exploring the timeline as a framework for meaning making}. In \bibinfo{booktitle}{\emph{Proceedings of the {SIGCHI} {Conference} on {Human} {Factors} in {Computing} {Systems}}}. \bibinfo{publisher}{ACM}, \bibinfo{address}{Paris France}, \bibinfo{pages}{1619--1628}.
\newblock
\showISBNx{978-1-4503-1899-0}
\urldef\tempurl%
\url{https://doi.org/10.1145/2470654.2466215}
\showDOI{\tempurl}


\bibitem[Thiry and Rosson(2012)]%
        {thiry_unearthing_2012}
\bibfield{author}{\bibinfo{person}{Elizabeth Thiry} {and} \bibinfo{person}{Mary~Beth Rosson}.} \bibinfo{year}{2012}\natexlab{}.
\newblock \showarticletitle{Unearthing the family gems: design requirements for a digital reminiscing system for older adults}. In \bibinfo{booktitle}{\emph{{CHI} '12 {Extended} {Abstracts} on {Human} {Factors} in {Computing} {Systems}}}. \bibinfo{publisher}{ACM}, \bibinfo{address}{Austin Texas USA}, \bibinfo{pages}{1715--1720}.
\newblock
\showISBNx{978-1-4503-1016-1}
\urldef\tempurl%
\url{https://doi.org/10.1145/2212776.2223698}
\showDOI{\tempurl}


\bibitem[Thudt et~al\mbox{.}(2016)]%
        {thudt_visual_2016}
\bibfield{author}{\bibinfo{person}{Alice Thudt}, \bibinfo{person}{Dominikus Baur}, \bibinfo{person}{Samuel Huron}, {and} \bibinfo{person}{Sheelagh Carpendale}.} \bibinfo{year}{2016}\natexlab{}.
\newblock \showarticletitle{Visual {Mementos}: {Reflecting} {Memories} with {Personal} {Data}}.
\newblock \bibinfo{journal}{\emph{IEEE Transactions on Visualization and Computer Graphics}} \bibinfo{volume}{22}, \bibinfo{number}{1} (\bibinfo{date}{Jan.} \bibinfo{year}{2016}), \bibinfo{pages}{369--378}.
\newblock
\showISSN{1077-2626, 1941-0506, 2160-9306}
\urldef\tempurl%
\url{https://doi.org/10.1109/TVCG.2015.2467831}
\showDOI{\tempurl}


\bibitem[Tokunaga et~al\mbox{.}(2021)]%
        {tokunaga_dialogue-based_2021}
\bibfield{author}{\bibinfo{person}{Seiki Tokunaga}, \bibinfo{person}{Kazuhiro Tamura}, {and} \bibinfo{person}{Mihoko Otake-Matsuura}.} \bibinfo{year}{2021}\natexlab{}.
\newblock \showarticletitle{A {Dialogue}-{Based} {System} with {Photo} and {Storytelling} for {Older} {Adults}: {Toward} {Daily} {Cognitive} {Training}}.
\newblock \bibinfo{journal}{\emph{Frontiers in Robotics and AI}}  \bibinfo{volume}{8} (\bibinfo{date}{June} \bibinfo{year}{2021}), \bibinfo{pages}{644964}.
\newblock
\showISSN{2296-9144}
\urldef\tempurl%
\url{https://doi.org/10.3389/frobt.2021.644964}
\showDOI{\tempurl}


\bibitem[Trafimow et~al\mbox{.}(1991)]%
        {trafimow_tests_1991}
\bibfield{author}{\bibinfo{person}{David Trafimow}, \bibinfo{person}{Harry~C. Triandis}, {and} \bibinfo{person}{Sharon~G. Goto}.} \bibinfo{year}{1991}\natexlab{}.
\newblock \showarticletitle{Some tests of the distinction between the private self and the collective self.}
\newblock \bibinfo{journal}{\emph{Journal of Personality and Social Psychology}} \bibinfo{volume}{60}, \bibinfo{number}{5} (\bibinfo{date}{May} \bibinfo{year}{1991}), \bibinfo{pages}{649--655}.
\newblock
\showISSN{1939-1315, 0022-3514}
\urldef\tempurl%
\url{https://doi.org/10.1037/0022-3514.60.5.649}
\showDOI{\tempurl}


\bibitem[Wang(2001)]%
        {wang_culture_2001}
\bibfield{author}{\bibinfo{person}{Qi Wang}.} \bibinfo{year}{2001}\natexlab{}.
\newblock \showarticletitle{Culture effects on adults' earliest childhood recollection and self-description: {Implications} for the relation between memory and the self.}
\newblock \bibinfo{journal}{\emph{Journal of Personality and Social Psychology}} \bibinfo{volume}{81}, \bibinfo{number}{2} (\bibinfo{year}{2001}), \bibinfo{pages}{220--233}.
\newblock
\showISSN{1939-1315, 0022-3514}
\urldef\tempurl%
\url{https://doi.org/10.1037/0022-3514.81.2.220}
\showDOI{\tempurl}


\bibitem[Wang(2008)]%
        {wang_cultural_2008}
\bibfield{author}{\bibinfo{person}{Qi Wang}.} \bibinfo{year}{2008}\natexlab{}.
\newblock \showarticletitle{On the cultural constitution of collective memory}.
\newblock \bibinfo{journal}{\emph{Memory}} \bibinfo{volume}{16}, \bibinfo{number}{3} (\bibinfo{date}{April} \bibinfo{year}{2008}), \bibinfo{pages}{305--317}.
\newblock
\showISSN{0965-8211, 1464-0686}
\urldef\tempurl%
\url{https://doi.org/10.1080/09658210701801467}
\showDOI{\tempurl}


\bibitem[Wang and Conway(2004)]%
        {wang_stories_2004}
\bibfield{author}{\bibinfo{person}{Qi Wang} {and} \bibinfo{person}{Martin~A. Conway}.} \bibinfo{year}{2004}\natexlab{}.
\newblock \showarticletitle{The {Stories} {We} {Keep}: {Autobiographical} {Memory} in {American} and {Chinese} {Middle}‐{Aged} {Adults}}.
\newblock \bibinfo{journal}{\emph{Journal of Personality}} \bibinfo{volume}{72}, \bibinfo{number}{5} (\bibinfo{date}{Oct.} \bibinfo{year}{2004}), \bibinfo{pages}{911--938}.
\newblock
\showISSN{0022-3506, 1467-6494}
\urldef\tempurl%
\url{https://doi.org/10.1111/j.0022-3506.2004.00285.x}
\showDOI{\tempurl}


\bibitem[Webster et~al\mbox{.}(2010)]%
        {webster_mapping_2010}
\bibfield{author}{\bibinfo{person}{Jeffrey~Dean Webster}, \bibinfo{person}{Ernst~T. Bohlmeijer}, {and} \bibinfo{person}{Gerben~J. Westerhof}.} \bibinfo{year}{2010}\natexlab{}.
\newblock \showarticletitle{Mapping the {Future} of {Reminiscence}: {A} {Conceptual} {Guide} for {Research} and {Practice}}.
\newblock \bibinfo{journal}{\emph{Research on Aging}} \bibinfo{volume}{32}, \bibinfo{number}{4} (\bibinfo{date}{July} \bibinfo{year}{2010}), \bibinfo{pages}{527--564}.
\newblock
\showISSN{0164-0275, 1552-7573}
\urldef\tempurl%
\url{https://doi.org/10.1177/0164027510364122}
\showDOI{\tempurl}


\bibitem[Westerhof and Bohlmeijer(2014)]%
        {westerhof_celebrating_2014}
\bibfield{author}{\bibinfo{person}{Gerben~J. Westerhof} {and} \bibinfo{person}{Ernst~T. Bohlmeijer}.} \bibinfo{year}{2014}\natexlab{}.
\newblock \showarticletitle{Celebrating fifty years of research and applications in reminiscence and life review: {State} of the art and new directions}.
\newblock \bibinfo{journal}{\emph{Journal of Aging Studies}}  \bibinfo{volume}{29} (\bibinfo{date}{April} \bibinfo{year}{2014}), \bibinfo{pages}{107--114}.
\newblock
\showISSN{08904065}
\urldef\tempurl%
\url{https://doi.org/10.1016/j.jaging.2014.02.003}
\showDOI{\tempurl}


\bibitem[{Wikipedia contributors}(2024)]%
        {wikipedia_contributors_retirement_2024}
\bibfield{author}{\bibinfo{person}{{Wikipedia contributors}}.} \bibinfo{year}{2024}\natexlab{}.
\newblock \bibinfo{title}{Retirement age}.
\newblock
\newblock
\urldef\tempurl%
\url{https://en.wikipedia.org/w/index.php?title=Retirement_age&oldid=1193931722}
\showURL{%
\tempurl}


\bibitem[Wildschut et~al\mbox{.}(2006)]%
        {wildschut_nostalgia_2006}
\bibfield{author}{\bibinfo{person}{Tim Wildschut}, \bibinfo{person}{Constantine Sedikides}, \bibinfo{person}{Jamie Arndt}, {and} \bibinfo{person}{Clay Routledge}.} \bibinfo{year}{2006}\natexlab{}.
\newblock \showarticletitle{Nostalgia: {Content}, triggers, functions.}
\newblock \bibinfo{journal}{\emph{Journal of Personality and Social Psychology}} \bibinfo{volume}{91}, \bibinfo{number}{5} (\bibinfo{date}{Nov.} \bibinfo{year}{2006}), \bibinfo{pages}{975--993}.
\newblock
\showISSN{1939-1315, 0022-3514}
\urldef\tempurl%
\url{https://doi.org/10.1037/0022-3514.91.5.975}
\showDOI{\tempurl}


\bibitem[Wu et~al\mbox{.}(2020)]%
        {wu_interactive_2020}
\bibfield{author}{\bibinfo{person}{Yi-Luen Wu}, \bibinfo{person}{Edwinn Gamborino}, {and} \bibinfo{person}{Li-Chen Fu}.} \bibinfo{year}{2020}\natexlab{}.
\newblock \showarticletitle{Interactive {Question}-{Posing} {System} for {Robot}-{Assisted} {Reminiscence} {From} {Personal} {Photographs}}.
\newblock \bibinfo{journal}{\emph{IEEE Transactions on Cognitive and Developmental Systems}} \bibinfo{volume}{12}, \bibinfo{number}{3} (\bibinfo{date}{Sept.} \bibinfo{year}{2020}), \bibinfo{pages}{439--450}.
\newblock
\showISSN{2379-8920, 2379-8939}
\urldef\tempurl%
\url{https://doi.org/10.1109/TCDS.2019.2917030}
\showDOI{\tempurl}


\bibitem[Xie et~al\mbox{.}(2012)]%
        {xie_connecting_2012}
\bibfield{author}{\bibinfo{person}{Bo Xie}, \bibinfo{person}{Allison Druin}, \bibinfo{person}{Jerry Fails}, \bibinfo{person}{Sheri Massey}, \bibinfo{person}{Evan Golub}, \bibinfo{person}{Sonia Franckel}, {and} \bibinfo{person}{Kiki Schneider}.} \bibinfo{year}{2012}\natexlab{}.
\newblock \showarticletitle{Connecting generations: developing co-design methods for older adults and children}.
\newblock \bibinfo{journal}{\emph{Behaviour \& Information Technology}} \bibinfo{volume}{31}, \bibinfo{number}{4} (\bibinfo{date}{April} \bibinfo{year}{2012}), \bibinfo{pages}{413--423}.
\newblock
\showISSN{0144-929X, 1362-3001}
\urldef\tempurl%
\url{https://doi.org/10.1080/01449291003793793}
\showDOI{\tempurl}


\bibitem[Xu et~al\mbox{.}(2023)]%
        {xu_effects_2023}
\bibfield{author}{\bibinfo{person}{Lijun Xu}, \bibinfo{person}{Shasha Li}, \bibinfo{person}{Renfu Yan}, \bibinfo{person}{Yingyuan Ni}, \bibinfo{person}{Yuecong Wang}, {and} \bibinfo{person}{Yue Li}.} \bibinfo{year}{2023}\natexlab{}.
\newblock \showarticletitle{Effects of reminiscence therapy on psychological outcome among older adults without obvious cognitive impairment: {A} systematic review and meta-analysis}.
\newblock \bibinfo{journal}{\emph{Frontiers in Psychiatry}}  \bibinfo{volume}{14} (\bibinfo{date}{March} \bibinfo{year}{2023}), \bibinfo{pages}{1139700}.
\newblock
\showISSN{1664-0640}
\urldef\tempurl%
\url{https://doi.org/10.3389/fpsyt.2023.1139700}
\showDOI{\tempurl}


\end{thebibliography}

\appendix
\section{pilot study details}
\label{appendix: pilot}
Before starting the user study, we conducted a pilot study with three Chinese older adults (gender: all females; age: median = 68, mean = 68.33, SD = 12.50) through online semi-structured interviews. The participants were recruited from our own networks and were asked to prepare their personal old photo collections to share with us. The main goal of this pilot study was to gain an initial understanding of older adults' photo collections, their use of technologies for photo-based reminiscence, and their preferences for designing supportive technologies. Each interview lasted around 30 minutes. After the interviews, three authors independently coded the transcripts by identifying key findings, such as "old photos convey important meanings," "connect stories to historical context," and "difficulty in imagining ideal display methods." These findings were then discussed in a 90-minute meeting.

The pilot study revealed that older adults have a deep passion for engaging with photos for reminiscence and exhibit certain preferences, such as their photo collections mainly featuring important people and a tendency to connect stories with historical context. Insights from these interviews, combined with existing literature, inspired the design of discussion cards for co-design activities in Section 3.2. We also confirmed that older adults use technologies, such as mobile phones, for various activities like sharing photos, supporting our decision to apply the PhotoUse model for a systematic understanding.

During the interviews, we attempted to get their opinions on how typical photo-related technologies, such as digital albums, could be improved based on their regular photo activities. However, we observed that older adults have difficulty articulating their preferences in detail. For example, when asked about displaying their digital photos and their preferred layout, they struggled to understand the concept of layout or to imagine possible designs. This led us to design discussion cards to help them better articulate their preferences. We incorporated insights from the interviews and existing literature, such as findings from Axtell et al.'s study \cite{axtell_design_2022}, to include familiar metaphors like digital album designs and various layout options. This approach enabled participants to compare and contrast different designs, thereby better articulating their preferences.

\section{definition of activities of the photouse model}
\label{appendix: model}
Reproduced from \cite{broekhuijsen_photowork_2017}

\textbf{Accumulating}

- Capturing: taking pictures; on-device quality triage to determine retaking the picture

- Collecting: adding pictures to your collection, which you did not capture yourself

\textbf{Curating}

- Organising: tagging, moving, categorising, naming, captioning, archiving, and deleting

- Triaging: assessing, selecting for a specific purpose (e.g. sharing, decorating, and presenting)

- Managing: filing, backup, downloading, and uploading

- Editing: retouching, cropping, combining, correcting, and changing

\textbf{Retrieving}

- Browsing: for example, browsing (casual viewing of pictures while interacting with them) 

- Viewing: passive viewing of slideshows

- Searching: for example, goal-directed retrieving, searching

\textbf{Appropriating}

- Sharing: remote sharing (online, on social media, or sending postcards), collocated sharing 

- Printing: printing photos, a poster, or family albums

- Collaging: making a collage from (printed) photos, making (digital) booklets

- Tinkering: tinkering with printed photos, cutting and pasting printed photos

\end{document}